\documentclass[a4paper,11pt]{article}
\usepackage{amssymb,amsmath}
\usepackage{mathrsfs}
\usepackage{xypic}
\usepackage{epic}

\allowdisplaybreaks
\numberwithin{equation}{section} 

\newtheorem{thm}{Theorem}[section]
\newtheorem{prop}[thm]{Proposition}
\newtheorem{lemma}[thm]{Lemma}
\newtheorem{defn}[thm]{Definition}

\newtheorem{example}[thm]{Example}

\newtheorem{remark}[thm]{Remark}

\newenvironment{rmk}{\begin{remark} \rm }{\end{remark}}
\newenvironment{prf}{{\it Proof }}{\hfill$\Box$}

\newcommand\od{\mathrm{d}}
\newcommand\ad{\mathrm{ad}}
\newcommand\tr{\mathrm{tr}}
\newcommand{\nn}{\nonumber}

\newcommand{\diag}{\mathrm{diag}}

\newcommand\pd{\partial}
\newcommand{\ld}{\lambda}
\newcommand{\al}{\alpha}

\newcommand{\Gm}{\Gamma}
\newcommand{\Ld}{\Lambda}
\newcommand{\Dt}{\Delta}
\newcommand{\dt}{\delta}

\newcommand{\ka}{\kappa}

\newcommand{\res}{\mathrm{res}}

\newcommand{\kn}{\mathrm{Ker}}
\newcommand{\im}{\mathrm{Im}}

\newcommand\C{\mathbb{C}}
\newcommand\R{\mathbb{R}}
\newcommand\Z{\mathbb{Z}}
\newcommand\Zop{\mathbb{Z^{\mathrm{odd}}_+}}

\newcommand\mA{\mathcal{A}}
\newcommand\mF{\mathcal{F}}
\newcommand\mD{\mathcal{D}}
\newcommand\mL{\mathcal{L}}
\newcommand\mR{\mathcal{R}}
\newcommand\msL{\mathscr{L}}
\newcommand\sP{\mathscr{P}}
\newcommand\sQ{\mathscr{Q}}
\newcommand\mfg{\mathfrak{g}}

\newcommand\ra{\rangle}
\newcommand\la{\langle}

\newcommand{\p}{\partial}

\begin{document}
\title{On the Drinfeld-Sokolov Hierarchies of D type}
\author{Si-Qi Liu\thanks{liusq@mail.tsinghua.edu.cn}\quad Chao-Zhong Wu\thanks{wucz05@mails.tsinghua.edu.cn}\quad Youjin
Zhang\thanks{youjin@mail.tsinghua.edu.cn} \\
{\small Department of Mathematical Sciences, Tsinghua University, }\\
{\small Beijing 100084, P. R. China}}
\date{}
\maketitle

\begin{abstract}
We extend the notion of pseudo-differential operators that are used
to represent the Gelfand-Dickey hierarchies, and obtain a similar
representation for the full Drinfeld-Sokolov hierarchies of $D_n$
type. By using such pseudo-differential operators we introduce the
tau functions of these bi-Hamiltonian hierarchies, and prove that
these hierarchies are equivalent to the integrable hierarchies
defined by Date-Jimbo-Kashiware-Miwa and Kac-Wakimoto from the basic
representation of the Kac-Moody algebra $D_n^{(1)}$.

\end{abstract}

\tableofcontents

\section{Introduction}
For every affine Lie algebra $\mfg$ and a choice of a vertex $c_m$
of the extended Dynkin diagram, Drinfeld and Sokolov constructed in
\cite{DS} a hierarchy of integrable systems which generalizes the
prototypical soliton equation--the Korteweg-de Vries equation. This
construction provides a big class of integrable hierarchies that are
important in different areas of mathematical physics. In particular,
the integrable hierarchies that are associated to the affine Lie
algebras of A-D-E type are shown to be closely related to 2d
topological field theory and Gromov-Witten invariants, see \cite{Du,
DZ, FSZ, FJR, Gi-A, GM, Kon, Witten1} and references therein. In
establishing such relationships the tau functions of the integrable
hierarchies play a crucial role, they correspond to the partition
functions of topological field theory models. The unknown functions
of the hierarchy are related to some special two point correlation
functions.

The  definition of the tau functions for the Drinfeld-Sokolov
hierarchies and their generalizations \cite{GHM} was given in
\cite{HM, HM-2} by using the dressing operators of the hierarchies.
In terms of the tau functions such integrable hierarchies and their
generalizations are represented as systems of Hirota bilinear
equations, they can also be constructed by using the representation
theoretical approach to solition equations developed by Date, Jimbo,
Kashiwara, Miwa~\cite{DKJM-KPBKP, DJKM-reduce} and by Kac,
Wakimoto~\cite{KW, Kac}. In this approach the systems of Hirota
bilinear equations are constructed from an integrable highest weight
representation of $\mfg$ and its vertex operator realization,  the
tau functions that satisfy these equations are elements of the orbit
of the highest weight vector of the representation under the action
of the affine Lie group. Note that tau functions of the
Drinfeld-Sokolov hierarchies are also defined in \cite{EF, Wi} via
certain symmetry (called tau-symmetry in \cite{DZ}) of the
Hamiltonian densities of the hierarchies represented in forms of
modified KdV type. Here the unknown functions of the
Drinfeld-Sokolov hierarchies in forms of modified KdV type and in
that of KdV type are related by Miura type transformations.

For general Drinfeld-Sokolov hierarchies there are no canonical
choices for their unknown functions, and the definition of the tau
functions given in \cite{EF, HM-2, HM} in terms of the dressing
operators is in certain sense implicit. However, in the particular
case when  the affine Lie algebra is $A_n^{(1)}$ the
Drinfeld-Sokolov hierarchy coincides with the Gelfand-Dickey
hierarchy \cite{GD76},  the unknown functions can be taken as the
coefficients of a differential operator
\[
L= D^{n+1}+u^{n} D^{n-1}+\ldots+u^{2} D+u^1, \quad D=\frac{\od}{\od
x},
\]
and the integrable hierarchy can be represented in the form
\begin{equation}\label{GD}
\frac{\pd L}{\pd t_k}=[(L^{\frac{k}{n+1}})_+, L], \quad
k=\Z_+\setminus (n+1)\Z_+.
\end{equation}
Here $u^i$ are functions of the spatial variable $x$ and the time variables $t_1,t_2,\dots$. This integrable hierarchy has the Hamiltonian structure
\[
\frac{\pd u^i}{\pd t_k}=\{u^i(x), H_{k+n+1}\},
\]
where the Poisson bracket is defined by
\[
\{F, G\}=\int  \res\left(\left[\frac{\delta F}{\delta L},L\right]\frac{\delta G}{\delta L}\right) \od x
\]
for local functionals $F$, $G$, and the densities of the
Hamiltonians $H_k=\int h_{k}(u, u_x,\dots) \od x$ can be chosen as
\[
h_k=\frac{n+1}{k} \res\,L^{\frac{k}{n+1}}.
\]
The advantage of such a choice of the Hamiltonian densities lies in the fact that they satisfy the tau symmetry condition
\[
\frac{k}{n+1}\frac{\p h_{k}}{\p t_{l}}
=\frac{l}{n+1}\frac{\p h_{l}}{\p t_{k}}.
\]
Due to this property of the densities the tau function of the Gelfand-Dickey hierarchy
can be introduced, as it was done in \cite{DKJM-KPBKP, DZ, FF,  Wi}, by the equations
\begin{equation}\label{zh-2}
\frac{\p^2 \log\tau}{\p x\p t_{k}}=\frac{k}{n+1} h_{k},\quad
k=\Z_+\setminus (n+1)\Z_+.
\end{equation}
Note that the Hamiltonians for the general Drinfeld-Sokolov hierarchies are also given in \cite{DS}, however the densities given there do not satisfy the tau symmetry condition. In order to fulfill such a condition these densities should be modified by adding certain terms which are total $x$-derivatives of some differential polynomials of the unknown functions.

In the above formalism of the Drinfeld-Sokolov hierarchy associated to the affine Lie algebra $A_n^{(1)}$,  the integrable hierarchy and the relation of its unknown functions with the tau function are relatively explicitly given.
The purpose of the present paper is to give a similar representation for the Drinfeld-Sokolov hierarchy associated to the affine Lie algebra $D_n^{(1)}$ and the vertex $c_0$ of the Dynkin diagram.  Such a formalism is helpful for people to have a clear picture of the relation of integrable systems with Gromov-Witten invariants and topological field models associated to A-D-E singularities \cite{FSZ, FJR, FGM, Gi-GW, Gi-A, GM, Wu}. In fact, Drinfeld and Sokolov already represented in \cite{DS} part of the integrable hierarchy in terms of a pseudo-differential operator
of the form
\begin{equation}
L=D^{2n-2}+\sum_{i=1}^{n-1}D^{-1}\left(u^i D^{2i-1}+D^{2i-1}u^i\right)+D^{-1}\rho D^{-1}\rho,\label{eq-LL}
\end{equation}
where the functions $u^1,\dots, u^{n-1}, u^n=\rho$ serve as the unknown functions of the hierarchy.
The integrable systems of the hierarchy can be labeled by the elements of a chosen base
\begin{equation}\label{base-s}
\{\Ld^j\in\mfg^j, \Gm^j\in\mfg^{j(n-1)}\mid j\in2\Z+1\}
\end{equation}
of the principal Heisenberg subalgebra of $D_n^{(1)}$ (see Sec.
\ref{sec-4-2} for the definition of these symbols). Denote by $P$
the fractional power $L^{\frac{1}{2n-2}}$ of $L$ which is a
pseudo-differential operator of the form
\[P=D+w_1 D^{-1}+w_2 D^{-2}+\dots,\]
then the part of the integrable hierarchy that corresponds to the elements $\Ld^j$ can be represented as \cite{DS}
\begin{equation}
\frac{\p L}{\p t_k}=[(P^k)_+, L],\quad k\in\Zop.
\end{equation}
The other part that corresponds to the elements $\Gamma^j$ can not be represented in this way by using only the pseudo-differential operators $L, P$.

Inspired by the Lax pair representations of the dispersionless integrable hierarchy that appear in 2D topological field theory \cite{Du,Ta},
we attempt to represent the flows corresponding to the elements $\Gm^j$ by the square root $Q$ of $L$ which takes the form
\[Q=D^{-1}\rho+\sum_{k\ge 0} w_k D^{k}.\]
However, this operator is not a pseudo-differential operator in the usual sense, because it contains infinitely many terms with positive powers of $D$,
so one cannot compute the square of $Q$. We note that in the dispersionless case, with $D$ replaced by its symbol $p$, one can define the square of
$Q$, and define the dispersionless hierarchy by using $L, P$ and $Q$.

We are to show in this paper that there exists a new kind of pseudo-differential operators which are allowed to contain infinitely many terms with positive
power of $D$ such as $Q$, so we can define the square root of the pseudo-differential operator $L$ in the space of such operators.
Then by using the pseudo-differential operators $L$ and $Q$ we can get the Lax pair representation of the remaining part
of the integrable hierarchy and define its tau function in a way
that one does for the Gelfand-Dickey hierarchy, see Theorem \ref{thm-main}. By using this new kind of pseudo-differential operators, we also find a Lax pair representation of
the two-component BKP hierarchy (see \cite{DJKM-KPtype}, c.f. \cite{Sh}). We show that the Drinfeld-Sokolov hierarchy of $D_n$ type becomes the
$(2n-2,2)$-reduction of the two-component BKP hierarchy \cite{DJKM-reduce}.
In this way we also prove that the square root of the tau function satisfies the Hirota bilinear equations that
are constructed in \cite{DJKM-reduce, KW} from the principal vertex
operator realization of the basic representation of the affine Lie
algebra $D_n^{(1)}$, see \eqref{bltau}, \eqref{tauhtau} and Theorem \ref{thm-bl}.

In order to obtain the above mentioned results, we first
extend, in Section 2, the usual definition of the ring of
pseudo-differential operators. Then in Section 3 we define a
hierarchy of integrable systems and its tau function by using the
pseudo-differential operator $L$ of the form \eqref{eq-LL} and its
fractional powers $P, Q$. In Section 4 we show that the constructed
hierarchy coincides with the Drinfeld-Sokolov hierarchy associated
to the affine Lie algebra $D_n^{(1)}$ and the vertex $c_0$ of its
Dynkin diagram. In Section 5 we give a Lax pair representation of the two-component BKP hierarchy, its tau function, and its $(2n-2,2)$-reductions.
In the final section we give some concluding remarks.

\section{Pseudo-differential operators}
In this section we generalize the concept of pseudo-differential
operators and list some useful properties of them.

\subsection{Definitions}

Let $\mA$ be a commutative ring with unity, and $D:\mA\to\mA$ be a
derivation. The algebra of pseudo-differential operators over $\mA$
is defined to be
\[\mD^-=\left\{\sum_{i < \infty}f_i D^i\mid f_i\in\mA\right\}.\]
This is a complete topological ring, whose topological basis is
given by the following filtration
\[\cdots\subset\mD^-_{(d-1)}\subset\mD^-_{(d)}\subset\mD^-_{(d+1)}\subset\cdots,\quad
 \mD^-_{(d)}=\left\{\sum_{i \le d}f_i D^i\mid f_i\in\mA\right\}.\]
The product of two pseudo-differential operators $A=\sum_{i\le k}f_i D^i\in \mD^-$ and $B=\sum_{j\le
l}g_j D^j\in \mD^-$ is defined by
\begin{equation}
A \cdot B= \sum_{i\le k}\sum_{j\le l} \sum_{r\ge0}\binom{i}{r}f_i
D^r(g_j)D^{i+j-r}\in\mD^-. \label{eq-prod}
\end{equation}
It is easy to see that for every $s\in\Z$, the coefficient of $D^s$
in \eqref{eq-prod} is a finite sum of elements of $\mA$, so the
above product is well defined.

In our formalism of the Drinfeld-Sokolov hierarchy of $D_n$ type below, one need not only operators in $\mD^-$ but also
operators in the following larger abelian group
\[\mD=\left\{\sum_{i \in \Z}f_i D^i\mid f_i\in\mA\right\}.\]
However, it is impossible to extend the product \eqref{eq-prod} to
$\mD$ because when expanding the product of two elements of $\mD$
one meets summations of infinitely many elements of $\mA$, which are
not well defined unless $\mA$ possesses certain topology.

Now we assume that on $\mA$ there is a gradation
\[\mA=\prod_{i\ge0}\mA_i, \quad \mA_i \cdot \mA_j \subset \mA_{i+j}\]
such that $\mA$ is topologically complete w.r.t. the induced decreasing filtration
\[\mA=\mA_0\supset\cdots\supset\mA_{(d-1)}\supset\mA_{(d)}\supset\mA_{(d+1)}\supset\cdots,\quad \mA_{(d)}=\prod_{i \ge d}\mA_i.\]

Let $D:\mA\to\mA$ be a derivation of degree one, i.e.
$D(\mA_i)\subset\mA_{i+1}$. An operator $A\in\mD^-\subset\mD$ is
said to be homogeneous if there exists an integer $k\in\Z$ such that
\[A=\sum_{i\le k}f_i D^i, \quad f_i\in\mA_{k-i},\]
and the integer $k$ is called the degree of $A$. We denote by
$\mD_k$ the subgroup that consists of all homogeneous
pseudo-differential operators of degree $k$, then the abelian group
$\mD$ has the following decomposition
\[\mD=\prod_{k\in\Z}\mD_k.\]
We introduce the following subgroups of $\mD$:
\[\mD^+_{(d)}=\prod_{k \ge d}\mD_k,\quad \mD^+=\bigcup_{d\in\Z}\mD^+_{(d)}.\]
It is easy to see that $\mD^+$ is topologically complete w.r.t. the
filtration
\[\cdots\supset\mD^+_{(d-1)}\supset\mD^+_{(d)}\supset\mD^+_{(d+1)}\supset\cdots.\]

For any $A\in\mD_k$ and $B\in\mD_l$, it is easy to see that their
product defined by \eqref{eq-prod} belongs to $\mD_{k+l}$, so we can
extend this product to $\mD^+$ such that $\mD^+$ becomes a ring.

\begin{defn}
Elements of $\mD^-$ (resp. $\mD^+$) are called pseudo-differential
operators of the first type (resp. the second type) over $\mA$. The
intersection of $\mD^-$ and $\mD^+$ in $\mD$ is denoted by
\[
\mD^b=\mD^-\cap \mD^+,
\]
and its elements are called  bounded pseudo-differential operators.

Sometimes to indicate the algebra $\mA$ and the derivation $D$, we
will use the notations $\mD^{\pm}(\mA, D)$ instead of $\mD^\pm$.
\end{defn}

The general form of $A\in\mD$ reads
\begin{equation}
A=\sum_{i\in\Z}\sum_{j\ge0}a_{i,j}D^i, \quad a_{i,j}\in \mA_j.
\label{eq-AA}
\end{equation}
The following lemma is obvious.
\begin{lemma} \label{lem-supp}
Suppose $A\in\mD$ is given in \eqref{eq-AA}, then
\begin{itemize}
\item[i)] $A\in\mD_k$ iff the coefficients $a_{i,j}$ are supported on the ray $\{(i,j)\mid
i+j=k, j\ge0\}$;
\item[ii)] $A\in\mD^+$ iff there exists $m\in\Z$
such that $a_{i,j}$ are supported on the domain $\{(i,j)\mid j\ge
\max\{0,m-i\}\}$;
\item[iii)] $A\in\mD^-$ iff there exists $n\in\Z$ such that $a_{i,j}$
are supported on the domain $\{(i,j)\mid i\le n, j\ge0\}$.
\end{itemize}
\end{lemma}
This lemma has a graphic interpretation as follows:
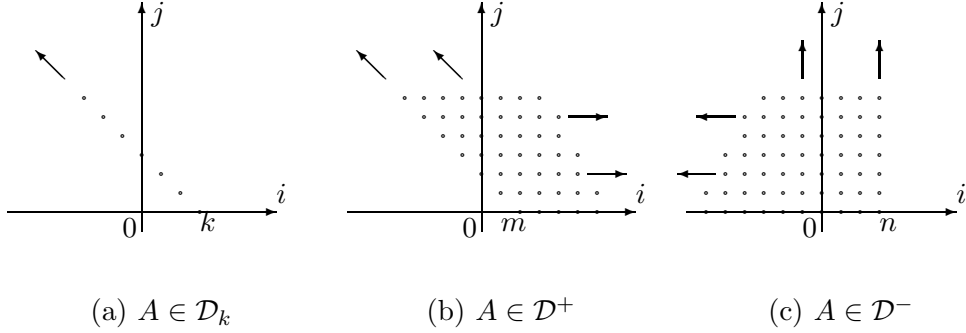
\[
\begin{array}{ccc}
\setlength{\unitlength}{0.02\textwidth}
\begin{picture}(16,12)
\put(0,0){\vector(1,0){14}} \put(7,-1){\vector(0,1){12}}
\put(14,0.5){$i$} \put(7.5,10){$j$} \put(6,-1.3){$0$}
\put(10,-1){$k$} \matrixput(10,0)(-1,1){7}(0,0){1}{\circle{0.2}}
\put(3,7){\vector(-1,1){1.5}}
\end{picture} &
\setlength{\unitlength}{0.02\textwidth}
\begin{picture}(16,12)
\put(0,0){\vector(1,0){15}} \put(7,-1){\vector(0,1){12}}
\put(15,0.5){$i$} \put(7.5,10){$j$} \put(6,-1.3){$0$}
\put(8,-1){$m$} \matrixput(9,0)(-1,1){7}(1,0){5}{\circle{0.2}}
\matrixput(13,1)(-1,1){6}(1,0){1}{\circle{0.2}}
\matrixput(12,3)(-1,1){4}(1,0){1}{\circle{0.2}}
\matrixput(11,5)(-1,1){2}(1,0){1}{\circle{0.2}}
\put(2,7){\vector(-1,1){1.5}} \put(6,7){\vector(-1,1){1.5}}
\put(12.5,2){\vector(1,0){2}} \put(11.5,5){\vector(1,0){2}}
\end{picture} &
\setlength{\unitlength}{0.02\textwidth}
\begin{picture}(16,12)
\put(0,0){\vector(1,0){14}} \put(7,-1){\vector(0,1){12}}
\put(14,0.5){$i$} \put(7.5,10){$j$} \put(6,-1.3){$0$}
\put(10,-1){$n$} \matrixput(10,0)(0,1){7}(-1,0){7}{\circle{0.2}}
\matrixput(3,0)(0,1){6}(-1,0){1}{\circle{0.2}}
\matrixput(2,0)(0,1){4}(-1,0){1}{\circle{0.2}}
\matrixput(1,0)(0,1){2}(-1,0){1}{\circle{0.2}}
 \put(10,7){\vector(0,1){2}} \put(6,7){\vector(0,1){2}}
  \put(1.5,2){\vector(-1,0){2}} \put(2.5,5){\vector(-1,0){2}}
\end{picture}   \\ \\
\mathrm{(a)}~A\in\mD_k & \mathrm{(b)}~A\in\mD^+ &
\mathrm{(c)}~A\in\mD^-
\end{array}
\]
From this interpretations it is easy to see the following alternative
expressions of the elements of $A\in\mD^{\pm}$.
\begin{itemize}
\item[i)] If $A\in\mD^+$, then there exists $m\in\Z$ and
$a_{i,j}\in\mA_j$ such that $A$ can be written as the following two
forms:
\begin{align}
&A=\sum_{i\in\Z}\left(\sum_{j\ge \max\{0,m-i\}}a_{i,j}\right)D^i, \label{dp-norm}\\
&A=\sum_{j\ge0}\left(\sum_{i\ge m-j}a_{i,j}D^i\right). \label{dp-disp}
\end{align}
\item[ii)] If $A\in\mD^-$, then there exists $n\in\Z$ and
$a_{i,j}\in\mA_j$ such that $A$ can be written as follows:
\begin{align}
&A=\sum_{i\le n}\left(\sum_{j\ge0}a_{i,j}\right)D^i, \label{dn-norm}\\
&A=\sum_{j\ge0}\left(\sum_{i\le n}a_{i,j}D^i\right). \label{dn-disp}
\end{align}
\end{itemize}
We call the expressions \eqref{dp-norm} and \eqref{dn-norm} the
\emph{normal expansion} of $A$, while the expressions
\eqref{dp-disp} and \eqref{dn-disp} the \emph{dispersion
expansion} of $A$.

Properties of pseudo-differential operators of the first type are
well known. Similar to the operators in $\mD^-$, we can define the
adjoint operator, the residue, the positive part and the negative
part of a pseudo-differential operator of the second type. Let
$A\in\mD^+$ be given by \eqref{eq-AA}, then
\begin{align*}
&A^*=\sum_{i\in\Z}\sum_{j\ge0}(-1)^iD^i\cdot a_{i,j},\quad \res\,A=\sum_{j\ge0}a_{-1,j},\\
&A_+=\sum_{i\ge0}\sum_{j\ge0} a_{i,j}D^i,\quad
A_-=\sum_{i<0}\sum_{j\ge0} a_{i,j}D^i.
\end{align*}
It is easy to see that $A^*, A_+, A_- \in \mD^+$ and
$\res\,A\in\mA$. In particular, if $A\in\mD^{\pm}$, then
$A_{\mp}\in\mD^b$.

An operator $A\in\mD^{\pm}$ is called a \emph{differential operator}
if its negative part $A_-$ vanishes. Note that every differential
operator in $\mD^-$ is of finite order, while the ones in $\mD^+$
may be not. The differential operators in $\mD^{\pm}$ form subrings
of $\mD^{\pm}$ respectively, and they can act on $\mA$ in the
obvious way. Given a differential operator $A\in\mD^{\pm}$, we
denote by $A(f)$ the action of $A$ on $f\in\mA$.

Let us introduce some other notations to be used latter. Elements of
the quotient space $\mF=\mA/D(\mA)$ are called \emph{local
functionals}, and they are represented in the form
\[\int\!\! f \od x=f+D(\mA),\quad f\in\mA.\]
Introduce the map
\[\la\, \,\ra:\ \mD\to\mF,\quad A\mapsto \la A\ra=\int \res A\, \od x.\]
We then define the pairing
\begin{equation}
\la A,B\ra=\la A B\ra\label{zh-11-20-1}
\end{equation}
on each of the following four spaces:
\[\mD^+\times\mD^+,\quad \mD^-\times\mD^-,\quad
\mD^b\times\mD,\quad \mD\times\mD^b.\] It is easy to see that this
pairing is symmetric and is nondegenerate on each of the above
spaces.

\subsection{Properties of pseudo-differential operators}

Now we present some useful properties of pseudo-differential
operators.

\begin{lemma}\label{lemma-fdo}
Let $A, B\in\mD^\pm$. If the commutator~$[A^m, B]=0$ for some positive integer~$m$, then~$[A, B]=0$.
\end{lemma}
\begin{prf}
The $\mD^-$ case is well known, we only prove the $\mD^+$ case.
Suppose $C=[A, B]\ne0$. We take the dispersion expansions
\[A=\sum_{j\ge a}\sum_{i\ge k_j} A_{i,j} D^i, \quad C=\sum_{j\ge c}\sum_{i\ge l_j} C_{i,j} D^i,\]
such that neither $A_{k_a,a}$  nor $C_{l_c,c}$ vanishes, then the
coefficient of~$D^{(m-1)k_a+l_c}$  in
\[[A^m, B]=[A, B] A^{m-1}+A [A,B] A^{m-2}+\cdots+ A^{m-1} [A,B]\]
reads
\[m A_{k_a,a}^{m-1} C_{l_c,c}+\cdots,\]
where $\cdots$ denote the terms with higher degrees in $\mA$. This
contradicts with~$[A^m, B]=0$. The lemma is proved.
\end{prf}

Let $\rho\in\mA$ be an invertible element, we consider the operator
\begin{equation}
Q=D^{-1}\rho+Q_+\in\mD^+, \label{eq-QQ}
\end{equation}
where $Q_+$ is a differential operator in $\mD^+$. Such an operator
$Q$ is invertible, whose inverse reads
\begin{align}
Q^{-1}=&\left(D^{-1}\rho(1+\rho^{-1}D Q_+)\right)^{-1} \nn \\
=&\left(1-\rho^{-1}D Q_++\rho^{-1}D Q_+\rho^{-1}D Q_+-\cdots\right)\rho^{-1}D. \label{Qinv}
\end{align}
Note that $Q^{-1}$ is a differential operator in $\mD^+$.

\begin{lemma}\label{thm-res}
Let $Q\in\mD^+$ be given in \eqref{eq-QQ}, then $D$ can be uniquely expressed as the following form
\begin{equation}\label{dq}
D=\sum_{i\ge 1}h_i Q^{-i}, \,\, h_i\in\mA.
\end{equation}
Moreover, $m\,h_m-\res\,Q^m \in D(\mA)$ for every $m\ge 1$.
\end{lemma}
\begin{prf} The first assertion follows from a simple induction.
We are going to prove the second one by using the following fact
\[\res\, Q^m= (D Q^m)_+ - D (Q^m)_+.\]
The first assertion shows that
\[(D Q^m)_+ =\left(\sum_{i\ge 1}h_i Q^{m-i}\right)_+ =\sum_{i\ge 1}h_i \left( Q^{m-i}\right)_+.\]
We assume $(Q^m)_+=\sum_{i\ge 0}a_{m, i} Q^{-i}$ with $a_{m, i}\in\mA$, then
\begin{align*}
D(Q^m)_+=&\sum_{i\ge0}a_{m, i}' Q^{-i}+\sum_{i\ge 0}a_{m, i}\sum_{j\ge 1} h_j Q^{-i-j} \\
=&\sum_{i\ge0}a_{m, i}' Q^{-i}+ \sum_{j\ge 1} h_j (Q^m)_+Q^{-j},
\end{align*}
where $a_{m, i}'=D(a_{m, i})$.

By using the above three formulae, one can obtain
\[\sum_{m\ge1} (\res\,Q^m)Q^{-m}=\sum_{i\ge1} \sum_{m=1-i}^0 h_i(Q^{m})_+Q^{-m-i}-\sum_{m\ge1}\sum_{i\ge0}a_{m,i}'Q^{-i-m}.\]
Note that $(Q^m)_+=Q^m$ when $m\le0$, so by comparing the
coefficients of $Q^{-m}$ we have
\[ m\,h_m-\res\,Q^m=\sum_{i=0}^{m-1}a_{m-i, i}'.\]
The lemma is proved.
\end{prf}

\begin{lemma} \label{lem-dec}
Let $A$ be a pseudo-differential operator in $\mD^+$, and
$\rho\in\mA$ be an invertible element. Then there exists a unique
pseudo-differential operator $B\in \mD^+$ such that $A=\rho B D+D B
\rho$. Furthermore, if $A^*=\pm A$, then $B^*=\mp B$.
\end{lemma}
\begin{prf}
Without loss of generality, we can assume $A$ to be homogeneous,
i.e., $A=\sum_{i\le k} a_i D^i$, $a_i\in\mA_{k-i}$. Suppose
$B=\sum_{i\le k-1} b_i D^i$, then one can determine $b_{k-1},
b_{k-2}, \dots$ recursively by  $A=\rho B D+D B \rho$. So we derive
the first part of the lemma.

If $A^*=\pm A$, then
\[\rho(B^*\pm B)D+D(B^*\pm B)\rho=0,\]
hence $B^*\pm B=0$ due to the uniqueness in the first part. The
lemma is proved.
\end{prf}

\section{An integrable hierarchy represented by pseudo-differential operators}

In this section we are to construct a hierarchy of evolutionary
partial differential equations starting from a pseudo-differential
operator $L$. This hierarchy possesses a bihamiltonian structure
which coincides with that of the Drinfeld-Sokolov hierarchy of
$D_n$ type, moreover, it possesses a tau function.

\subsection{Construction of the hierarchy}

Let $M$ be an open ball of dimension $n$ with coordinates $(u^1,
u^2, \dots, u^n)$. We define the algebra $\mA$ of differential
polynomials on $M$ to be
\[\mA=C^{\infty}(M)[[u^{i,s}\mid i=1, \dots, n,\ s=1, 2, \dots]].\]
There is a gradation on $\mA$ defined by
\[\deg f=0\mbox{ for } f\in C^\infty(M),\quad \deg u^{i,s}=s,\]
then it is easy to see that $\mA$ is topologically complete. We
introduce a derivation $D$ of degree one over $\mA$ as follows
\[D:\mA\to\mA, \quad D=\sum_{s\ge0}\sum_{i=1}^nu^{i,s+1}\frac{\p}{\p u^{i,s}},\]
where $u^{i,0}=u^i$. Now let us construct the algebras $\mD^{\pm}$
starting from $\mA$ and $D$ as we did in the last section.

Let $L$ be the following pseudo-differential operator given in \eqref{eq-LL}.
Obviously $L$ belongs to $\mD^b=\mD^-\cap\mD^+$ and satisfies
$L^*=DLD^{-1}$. Here we re-denote the coordinate $u^n$ by $\rho$,
and will use this notation frequently in what follows.

Firstly, we regard $L$ as an element of  $\mD^-$, then by using properties of the usual
pseudo-differential operators we have the following lemma.
\begin{lemma}\label{lem-3-1}
There exists a unique pseudo-differential operator $P\in\mD^-$ of
the form
\begin{equation}
P=D+u_1D^{-1}+u_2D^{-2}+\cdots \label{eq-P}
\end{equation}
such that $P^{2n-2}=L$. Moreover, the operator $P$ satisfies $[P,
L]=0$ and
\begin{equation}\label{Pstar}
P^*=-DPD^{-1}.
\end{equation}
\end{lemma}
In \cite{DKJM-KPBKP}, Date, Jimbo, Kashiwara and Miwa proved the
following lemma.
\begin{lemma}[\cite{DKJM-KPBKP}]\label{lem-3-2}
The constraint \eqref{Pstar} to an operator $P$ of the
form~\eqref{eq-P} is equivalent to the condition  that for every $k\in\Zop$ the
free term of $(P^k)_+$ vanishes, i.e. $(P^k)_+ (1)=0$.
\end{lemma}

The above two lemmas imply that the following equations
\begin{equation}\label{dn1}
\frac{\p L}{\p t_k}=[(P^k)_+, L],\quad k\in\Zop
\end{equation}
are well defined, and they give evolutionary partial differential
equations of $u^1,\dots u^n$. In particular, $D=\frac{\od}{\od x}$ with
$x=t_1$, and by taking residue of $D\left(\frac{\p L}{\p
t_k}-[(P^k)_+, L]\right)$ one has
\begin{equation}\label{rhot}
\frac{\pd \rho}{\pd t_k}=-(P^k)_+^*(\rho).
\end{equation}
The flows in \eqref{dn1} first appeared in \cite{DS} as part of the
Drinfeld-Sokolov hierarchy of $D_n$ type.

Note that the Drinfeld-Sokolov hierarchy of $D_n$ type contains $n$
series of commuting flows, but there are only $n-1$ series of flows
given in \eqref{dn1}, so in this sense the equations \eqref{dn1} do
not form a complete integrable hierarchy. One main result in the
present paper is that the $n$th series of flows of the
Drinfeld-Sokolov hierarchy of $D_n$ type can be represented by the
square root of $L$ regarded as an element of $\mD^+$.

\begin{lemma}\label{lemma-lq}
There exists a unique pseudo-differential operator $Q\in\mD^+$ of the following form
\begin{equation}\label{q}
Q=D^{-1}\rho+\sum_{m\ge0}Q_m\,D
\end{equation}
such that $Q^2=L$. Here $Q_m$ are homogeneous differential operators in $\mD^b$ with degree $2\,m$, and satisfy $Q_m^*=Q_m$.
Moreover, the operator $Q$ satisfies
\begin{align} \label{Qstar}
&Q^*=-D Q  D^{-1},\\
-&Q^*_+(\rho)=\frac{1}{2}D L_+(1). \label{prho}
\end{align}
\end{lemma}
\begin{prf}
By substituting \eqref{eq-LL} and \eqref{q} into $D Q^2=D L$ and
comparing the homogeneous terms, we can obtain
\begin{equation}
\rho Q_{m} D+D Q_{m}\rho =A_m,\quad m=0, 1, 2, \dots. \label{expan}
\end{equation}
Here $A_m$ are differential operators depending on
$L,Q_0,Q_1,\dots,Q_{m-1}$ and satisfy $A_m+A_m^*=0$. Then according
to Lemma \ref{lem-dec}, $Q_m$ can be determined by induction, and
they satisfy $Q_m^*=Q_m$.

The symmetry property \eqref{Qstar} is trivial. To show
\eqref{prho}, we consider the free terms on both hand sides of
\eqref{expan}:
\[ D Q_m(\rho)=\left\{\begin{array}{cl} u^{m+1,2m+1}, & m=0, 1,\ldots, n-2,
\\ 0, & m\ge n-1. \end{array}\right.\]
Hence
\[-Q^*_+(\rho)=\sum_{m\ge0}D Q_m(\rho)=\sum_{m=0}^{n-2}u^{m+1,2m+1}=\frac{1}{2} D L_+(1).\]
The lemma is proved.
\end{prf}

According to Lemmas \ref{lemma-fdo} and \ref{lemma-lq}, the
following evolutionary equations are well defined:
\begin{equation}\label{dn2}
\frac{\pd L}{\pd \hat{t}_{k}}=[-(Q^k)_-,L]=[(Q^k)_+,L], \quad
k\in\Zop.
\end{equation}
In particular, we have
\begin{equation}\label{rhoth}
\frac{\pd\rho}{\pd \hat{t}_k}=-(Q^k)^*_+(\rho), ~~k\in\Zop.
\end{equation}
Whe $k=1$ we obtain ${\pd\rho}/{\pd \hat{t}_1}=\frac{1}{2} D L_+(1)$, this flow
is linearly independent with ${\p \rho}/{\p t_{2i-1}}$ ($1\le i\le
n-1$), so from the bihamiltonian recursion relation (see
below) we see that the equations given in \eqref{dn1} are linearly independent with
that defined in \eqref{dn2}.
\begin{thm}
The flows in~\eqref{dn1}, \eqref{dn2} commute with each other.
\end{thm}
\begin{prf}
The commutativity of these flows follows from the following
equivalent representations of \eqref{dn1}, \eqref{dn2}:
\begin{align}\label{Pt}
&\frac{\pd P}{\pd {t}_k}=[(P^k)_+, P], \quad \frac{\pd P}{\pd
\hat{t}_k}=[-(Q^k)_-, P],\\
&\frac{\pd Q}{\pd t_k}=[(P^k)_+, Q], \quad \frac{\pd Q}{\pd
\hat{t}_k}=[-(Q^k)_-, Q], \label{Qt}
\end{align}
which can be verified as Lemma~\ref{lemma-fdo}. The theorem is
proved.
\end{prf}

The dispersionless limit of the flows $\frac{\pd}{\pd\hat{t}_k}$ was
first given by Takasaki in~\cite{Ta}, but the dispersionful one was not  given there.
Following~\cite{Ta}, we call the flows \eqref{dn1} and
\eqref{dn2} the \emph{positive} and the
\emph{negative} flows respectively. The above theorem shows that the negative and
the positive flows form an integrable hierarchy. We will show that it is
equivalent to the Drinfeld-Sokolov hierarchy of $D_n$ type.

\subsection{Bihamiltonian structure and tau structure}
In this subsection we show that the hierarchy \eqref{dn1},
\eqref{dn2} carries a bihamiltonian structure, and the densities of the Hamiltonians can be chosen to satisfy the tau symmetry condition. We then define the tau function of the hierarchy by using this tau symmetry following the
approach of \cite{DZ}.

Let  $\mL=D L$, it has the form
\begin{equation}\label{mL}
   \mL=D^{2n-1}+\sum_{i=1}^{n-1} \left(u^i
D^{2i-1}+D^{2i-1} u^i\right) + \rho D^{-1} \rho.
\end{equation}
Given a local functional $F=\int f\,\od x\in\mA/D(\mA)$,  we define
its variational derivative  w.r.t. $\mL$ to be an element $X={\dt
F}/{\dt\mL}\in\mD$ such that
\begin{equation}\label{zh-11-20}
\delta F=\la X, \delta \mL \ra, \quad X=X^*.
\end{equation}
The existence of such an element can be verified by taking
\begin{equation}\label{zh-11-1}
X=\frac1{2}\sum_{i=0}^{n-1}\left(D^{-2i}\frac{\dt F}{\dt
v^i(x)}+\frac{\dt F}{\dt v^i(x)}D^{-2i}\right).
\end{equation}
where $v^0=\rho^2$ and $v^1,\dots,v^{n-1}$ are determined by
representing the operator $\mL$ in the following form
\[
\mL=D^{2n-1}+\sum_{i=1}^{n-1}v^i
D^{2i-1}+\sum_{i=1}^{n-1}\tilde{v}^i D^{2i-2}+\rho D^{-1} \rho.
\]
Note that the new coordinates $v^1,\dots, v^{n-1}$ are related to
$u^1,\dots, u^{n-1}$ by a Miura-type transformation, and the
functions $\tilde{v}^i$ determined by the condition $\mL+\mL^*=0$
are linear functions of the derivatives of $v^1,\dots, v^{n-1}$.

On the other hand, the variational derivative $X$ defined in
\eqref{zh-11-20} is determined up to the addition of a kernel part
$Z$ that satisfies
\[Z_+(\rho)=0, \quad Z_-=\sum_{i\le n} \left(w_iD^{-2i}+D^{-2i}w_i\right), ~~w_i\in\mA.\]

The following compatible Poisson brackets are given in  Proposition
8.3 of \cite{DS} (see also \cite{DLZ}) for the bihamiltonian
structure of the Drinfeld-Sokolov hierarchy of $D_n$ type:
\begin{align}\label{poi1}
 \{F, G \}_1(\mathcal{L})&=\la X,  ( D Y_+\mL)_- - (\mL Y_+  D)_-
+ (\mL Y_- D)_+ - ( D Y_-\mL)_+   \ra, \\
\{F, G \}_2(\mathcal{L})&=\la X,
 (\mathcal{L} Y)_+ \mathcal{L} -  \mathcal{L}(Y \mathcal{L})_+
\ra, \label{poi2}
\end{align}
where $F$ and $G$ are two arbitrary local functionals, and
\[X=\frac{\dt F}{\dt\mL},\quad Y=\frac{\dt G}{\dt \mL}.\]
Note that in the above formulae of the Poisson brackets the second
component in the pairing $\la\,,\,\ra$ belongs to $\mD^b$ for any
$Y\in\mD$, so from the definition of $\la\,,\,\ra$ given in
\eqref{zh-11-20-1} we see that the first component $X$ is not
restricted to the space $\mD^+$ or $\mD^-$. One can show by a direct
computation that the definition of these Poisson brackets is
independent of the choice of the kernel parts of $X$ and $Y$, so
they are well defined.

\begin{thm} \label{thm-bh}
The hierarchy \eqref{dn1}, \eqref{dn2} has the following
bihamiltonian representation:
\begin{align}\label{Ft}
&\frac{\pd F}{\pd t_k}=\{F, H_{k+2n-2}\}_1=\{F, H_k\}_2,\\
&\frac{\pd F}{\pd\hat{t}_k}=\{F, \hat{H}_{k+2}\}_1=\{F,
\hat{H}_k\}_2.\label{Ft-b}
\end{align}
Here $F\in\mF$ is any local functional, and  the Hamiltonians are
given by
\begin{equation}\label{HkHhk}
H_k=\frac{2n-2}{k}\la P^k \ra, ~~
\hat{H}_{k}=\frac{2}{k}\la Q^k \ra,\quad k\in\Zop.
\end{equation}
\end{thm}
\begin{prf}
Let us start with the computation of the variational derivatives of the Hamiltonians
$H_k$. By using
the identity $P^{2n-2}=L$ (see Lemma \ref{lem-3-1}) and the symmetric property of the pairing $\la\,,\,\ra$ we have
\begin{align}
\dt H_k&=(2n-2)\la P^{k-1} , \dt P\ra=(2n-2)\la P^{k-2n+2} , P^{2n-3}\dt P\ra\nn\\
&=\la P^{k-2n+2} , \dt L\ra=\la P^{k-2n+2} D^{-1}, \dt\mL\ra
=\la Y_k, \dt\mL\ra,
\end{align}
where $Y_k=P^{k-2n+2} D^{-1}\in\mD$.  From \eqref{Pstar} it follows
that $Y_k^*=Y_k$, so we can take
\begin{equation}\label{zh-11-20-2}
\frac{\dt H_k}{\dt \mL}=Y_k=P^{k-2n+2} D^{-1}.
\end{equation}

To show~\eqref{Ft}, we first note due to Lemma \ref{lem-3-2} the validity of
\begin{align}
&D (P^k)_+ D^{-1}=D(P^k D^{-1})_+=(D P^k D^{-1})_+,\nn\\
&(P^k D^{-1} D)_-=(P^k D^{-1})_- D
\end{align}
for any $k\in\Zop$.
So from \eqref{dn1} we have
\begin{align*}
\frac{\pd \mL}{\pd t_k}&=D (P^k)_+ L-D L (P^k)_+
=(D P^k D^{-1})_+ \mL-\mL (P^k)_+\\
&=(\mL Y_k)_+\mL-\mL (Y_k \mL)_+.
\end{align*}
On the other hand, by using the commutativity between $L$ and $P$ (see Lemma \ref{lem-3-1})
we can also represent $\frac{\pd \mL}{\pd t_k}$ in the following form:
\begin{align*}
\frac{\pd \mL}{\pd t_k}&=D (P^k)_+ L-D L (P^k)_+\nn\\
&=\left(D (P^k)_+ L-D L (P^k)_+\right)_++\left(D (P^k)_+ L-D L (P^k)_+\right)_-\nn\\
&=\left(-D (P^k)_- L+D L (P^k)_-\right)_++\left(D (P^k)_+ L-D L (P^k)_+\right)_-\nn\\
&=\left(\mL (Y_{k+2n-2})_-D-D (Y_{k+2n-2})_-\mL\right)_+\nn\\
&\quad+\left(D (Y_{k+2n-2})_+\mL-\mL (Y_{k+2n-2})_+ D\right)_-
\end{align*}
Now the equivalence of the flows \eqref{dn1} with \eqref{Ft} follows from the above identities together with the relation
\[\frac{\pd F}{\pd t_k}=\left\la\frac{\dt F}{\dt \mL}, \frac{\pd \mL}{\pd t_k}\right\ra.\]

By using the property \eqref{Qstar} of the operator $Q$ we know that
for any $k\in\Zop$ the free term of $Q^k$ vanishes, then a similar
argument as above leads to the equivalence of the flows \eqref{dn2}
with \eqref{Ft-b}. The theorem is proved.
\end{prf}

By using the formula \eqref{zh-11-20-2} and
\[\frac{\dt \hat{H}_k}{\dt \mL}=Q^{k-2} D^{-1},\]
we obtain the following proposition.
\begin{prop}
The local functionals~$H_1, H_3,\ldots, H_{2n-3}$ and~$\hat{H}_1$ are
linearly independent Casimirs of the first Poisson bracket~$\{\, ,
\}_1$.
\end{prop}

We now verify that the above defined densities of the Hamiltonians satisfy the tau symmetry condition, and we can thus define the tau function for the integrable hierarchy \eqref{dn1}, \eqref{dn2}.
To this end let us introduce a series of rescaled time variables
\[T^{\al,p}=\left\{
\begin{array}{cl}
\dfrac{(2n-2)\Gm(p+1+\frac{2\al-1}{2n-2})}{\Gm(\frac{2\al-1}{2n-2})}t_{(2n-2)p+2\al-1},
& \al=1, \dots, n-1, \\ \\
\dfrac{2\Gm(p+1+\frac12)}{\Gm(\frac12)}\hat{t}_{2p+1}, & \al=n
\end{array}\right.\]
with $p=0, 1, 2, \dots$. Then the Hamiltonian equations \eqref{Ft},
\eqref{Ft-b} read
\[\frac{\p F}{\p T^{\al, p}}=\{F, H_{\al, p}\}_1=\left(p+\frac12+\mu_\al\right)^{-1}\{F, H_{\al, p-1}\}_2,\]
where the densities of the Hamiltonians $H_{\al, p}$ are given by
\[h_{\al,p-1}=\left\{
\begin{array}{cl}
\dfrac{\Gm(\frac{2\al-1}{2n-2})}{(2n-2)\,\Gm(p+1+\frac{2\al-1}{2n-2})}\res\,P^{(2n-2)p+2\al-1}, & \al=1, \dots, n-1, \\ \\
\dfrac{\Gm(\frac12)}{2\,\Gm(p+1+\frac12)}\res\,Q^{2p+1}, & \al=n,
\end{array}\right.\]
and the constants $\mu_\al$ are the spectrum of the underlying
Frobenius manifold \cite{Du, DLZ}, read
\[\mu_\al=\left\{
\begin{array}{cl}
\dfrac{2\al-n}{2n-2}, & \al=1, \dots, n-1, \\
0, & \al=n.
\end{array}\right.\]
Then we have tau symmetry
\[
\frac{\p h_{\al,p-1}}{\p T^{\beta,q}}=\frac{\p h_{\beta,q-1}}{\p T^{\al,p}},
\]
and the differential polynomials
\[
\Omega_{\al,p;\beta,q}=\p_x^{-1}\frac{\p h_{\al, p-1}}{\p
T^{\beta,q}}, \quad \al, \beta=1, 2, \dots, n; ~~p, q\ge0.
\]
have the property
\[ \Omega_{\al,p;\beta,q}= \Omega_{\beta,q;\al,p}.
\]
Hence the chosen $h_{\al, p}$ give a tau structure, in the sense of \cite{DZ}, of the
bihamiltonian structure of the integrable hierarchy \eqref{dn1},
\eqref{dn2}. This tau
structure defines the tau function $\hat{\tau}$ of the integrable hierarchy by
\begin{equation}\label{tauT}
\frac{\p^2 \log \hat{\tau}}{\p T^{\al,p}\p
T^{\beta,q}}=\Omega_{\al,p;\beta,q}.
\end{equation}

\section{Drinfeld-Sokolov hierarchies and pseudo-differential operators}\label{sec-4-2}
In this section we first recall some facts about the
Drinfeld-Sokolov hierarchies associated to untwisted affine Lie
algebras, see details in \cite{DS}. Then we consider the
Drinfeld-Sokolov hierarchy of $D_n$ type and identify it with the
hierarchy \eqref{dn1}, \eqref{dn2} constructed in the last section.

\subsection{Definition of the Drinfeld-Sokolov hierarchies}
Let $\mfg$ be an untwisted affine Lie algebra, and $\{ e_i, f_i,
h_i\mid i=0, 1, 2, \ldots, n\}$ be a set of Weyl generators of
$\mfg$. In Drinfeld and Sokolov's construction, the central element
$c$ is not used, so we always assume $c=0$. We need to use the
following two gradations on $\mfg$ \cite{DS, Kac}:
\begin{itemize}
\item[i)]  the principal/canonical gradation
\[\mfg=\bigoplus_{j\in\Z}\mfg^j,\quad \deg{{e}_i}=-\deg{{f}_i}=1,
\quad i=0, 1, \dots, n;\]
\item[ii)]  the homogeneous/standard gradation
\[\mfg=\bigoplus_{j\in\Z}\mfg_j, \quad \deg{{e}_i}=-\deg{{f}_i}=\dt_{i0},
\quad i=0, 1, \dots, n.\]
\end{itemize}
We will use notations such as~$\mfg^{<0}=\sum_{i<0}\mfg^i$
below.

In \cite{DS} Drinfeld and Sokolov assigned a standard
gradation to any chosen vertex $c_i$ of the Dynkin diagram of $\mfg$
and used the standard gradation to construct an integrable
hierarchy. As mentioned in the beginning of the present paper, we
only consider the case that the vertex is chosen to be $c_0$ which
is the special one added to the Dynkin diagram of the corresponding
simple Lie algebra. Integrable hierarchies that associated to different choices of
the vertices are related by Miura type transformations.

Denote by $E$ (resp. $E_+$) the set of exponents (resp. positive
exponents) of~$\mfg$.  Let $\mathfrak{s}$ be the Heisenberg
subalgebra associated to the principal gradation, which is defined
to be the centralizer of $\Ld=\sum_{i=0}^n e_i$. One can fix a basis
$\ld_j\in\mfg^j\ (j \in E)$ of $\mathfrak{s}$.

Let $C^\infty(\R,W)$ be the set of smooth functions from $\R$ to a linear space $W$.
We consider operators of the form
\begin{equation}\label{msL}
\msL= D+\Ld + q, \quad q\in C^\infty(\R, \mfg_0\cap\mfg^{\le0}),
\end{equation}
where $D=\frac{\od}{\od x}$, and $x$ is the coordinate on $\R$.
\begin{prop}[\cite{DS}]\label{prop-drg}
There exists an element~$U\in C^\infty(\R, \mfg^{< 0})$ such that
the operator $\msL_0=e^{-\ad_U}\msL$ has the form
\begin{equation}\label{L0g}
\msL_0= D+\Ld+H, \quad H\in C^\infty(\R, \mathfrak{s}\cap\mfg^{<0}),
\end{equation}
and for different choices of $U$, the map $H$ differs by the
addition of the total derivative of a differential polynomial of
$q$.
\end{prop}

We fix a $U$ as given in the above proposition, and introduce a map
\begin{equation}
\varphi:C^\infty(\R, \mfg)\to C^\infty(\R, \mfg), \quad A\mapsto
e^{\ad_U} A. \label{phi}
\end{equation}
The Drinfeld-Sokolov hierarchy is a hierarchy of partial
differential equations of gauge equivalence classes of $\msL$
defined by
\begin{equation}\label{dnm}
\frac{\pd \msL}{\pd t_j}=[\varphi(\ld_j)^+, \msL],  \quad j\in E_+.
\end{equation}
Here $\varphi(\ld_j)^+$ stands for the projection of
$\varphi(\ld_j)$ onto $C^\infty(\R,\mfg^{>0})$, and the gauge
transformations of $\msL$ read
\begin{equation}\label{gauge}
\msL\mapsto e^{\ad_N}\msL,  \quad N\in C^\infty(\R, \mfg_0\cap\mfg^{<0}).
\end{equation}

\begin{thm}[\cite{DS}]\label{th-ds}
The  Drinfeld-Sokolov hierarchy carries a bihamiltonian structure,
and the Hamiltonian densities are given by the expansion
coefficients of the map $H$ \eqref{L0g} in the basis~$\{\ld_{-j}\mid
j\in E_+\}$.
\end{thm}

For the classical untwisted affine Lie algebras, Drinfeld and
Sokolov proposed a way to represent their hierarchies via certain
scalar pseudo-differential operators over $\mA$, the algebra of
gauge invariant differential polynomials of $q$ in \eqref{msL}. They
gave such representations for the full hierarchies of the
$A_n^{(1)}$, $B_n^{(1)}$, $C_n^{(1)}$ types by using
pseudo-differential operators of the first type. However, for the
$D_n^{(1)}$ case, as pointed out by Drinfeld and Sokolov, the
pseudo-differential operators in $\mD^-$ are not enough to represent
the full hierarchy. Our purpose of introducing the space $\mD^+$ in
the present paper is to represent the full Drinfeld-Sokolov
hierarchy of $D_n$ type in terms of scalar pseudo-differential
operators.

The following lemma tells how to construct scalar
pseudo-differential operators from the operator $\msL$.
\begin{lemma}[\cite{DS}]\label{lemma-Dt}
Let~$\mathcal{R}$ be a ring with unity. We consider matrices of the
form
\[
R=\left(
    \begin{array}{cc}
      \al^t & a \\
      R_1 & \beta \\
    \end{array}
  \right)\in
\mathcal{R}^{m\times m},
\]
in which the block $R_1\in\mathcal{R}^{(m-1)\times(m-1)}$ is
invertible, $\al$, $\beta$ are $(m-1)$-dimensional column vectors,
and the superscript $t$ means the transpose of matrices. Define
$\Dt(R)=a-\al^t R_1^{-1}\beta$,  then the following statements are
true.
\begin{itemize}
\item[i)] Suppose~$x_1, x_2,\ldots, x_m, y$ belong to
some~$\mathcal{R}$-module such that
 \[
R\cdot(x_1, x_2,\ldots, x_m)^t=(y, 0,\ldots, 0)^t,
 \]
then $\Dt(R)\cdot x_m=y. $
\item[ii)] For any upper triangular matrix~$\tilde{N}\in
\mathcal{R}^{m\times m}$ with unity on the main diagonal one has
$\Dt(\tilde{N}R\tilde{N}^{-1})=\Dt(R)$.
\item[iii)] Given an anti-isomorphism $*$ of $\mathcal{R}$, one can define an
anti-isomorphism $T$  of $\mathcal{R}^{m\times m}$ by
$(R^T)_{ij}=R_{m+1-j, m+1-i}^*$. It satisfies $\Dt(R^T)=\Dt(R)^*$.
\end{itemize}
\end{lemma}

\subsection{Positive flows of the Drinfeld-Sokolov hierarchy of $D_n$
type}\label{sec-42}
In this subsection, we recall the approach given in
\cite{DS} that represents part of the Drinfeld-Sokolov hierarchy
of $D_n$ type as the positive flows \eqref{dn1} by using pseudo-diferential operators.

We first recall the matrix realization of the affine Lie algebra
$\mfg$ of $D_n^{(1)}$ type \cite{Kac, DS}. Denote by~$e_{i,j}$
the~$2n\times 2n$ matrix that takes value $1$ at the~$(i,j)$-entry
and zero elsewhere, then one can realize~$\mfg$ by choosing the Weyl
generators as follows:
\begin{align}\label{efhh}
&e_0=\frac{\ld}{2}(e_{1,2n-1}+e_{2,
2n}),~e_n=\frac{1}{2}(e_{n+1,n-1}+e_{n+2,n}),\\
&e_i=e_{i+1,i}+e_{2n+1-i,2n-i}~ (1\le i\le n-1),  \\
&f_0=\frac{2}{\ld}(e_{2n-1,1}+e_{2n,2}),~f_n={2}(e_{n-1,n+1}+e_{n,n+2}),\\
&f_i=e_{i,i+1}+e_{2n-i,2n+1-i}~ (1\le i\le n-1),\\
&h_i=[e_i,f_i] ~(0\le i\le n).
 \label{efht}
\end{align}
In particular, the associated simple Lie algebra $\mfg_0$ of $D_n$ type is
realized as
\begin{equation}\label{g}
\mfg_0=\Big\{A\in\C^{2n\times 2n}\mid A=-S A^T  S^{-1}\Big\},
\end{equation}
where $S$ is the following matrix
\[S=\sum_{i=1}^{n}(-1)^{i-1}(e_{i,i}+e_{2n+1-i,2n+1-i}),\]
and $A^T=(a_{l+1-j,k+1-i})$ for any~$k\times l$ matrix $A=(a_{ij})$.
Note that in this realization the algebra $\mfg$ is just
$\mfg_0\otimes \C[\lambda, \lambda^{-1}]$.

The set of exponents of~$\mfg$ is given by
\[
E=\{1, 3, 5,\ldots,2n-3\}\cup\{(n-1)'\}+(2n-2)\Z,
\]
where $(n-1)'$ indicates that when $n$ is even the multiplicity of
each exponent congruent to $n-1$ modulo $2n-2$ is $2$. A basis of
the principal Heisenberg subalgebra~$\mathfrak{s}$ can be chosen as
\[
\{-\Ld^k\in\mfg^k, \Gm^k\in\mfg^{k(n-1)}\mid k\in2\Z+1\},\label{zh-12-29a}
\]
where $\Ld=\sum_{i=0}^{n}{e}_i$, and
\begin{align}
\Gm=& \kappa\Big(
e_{n,1}-\frac{1}{2}e_{n+1,1}-\frac{\ld}{2}e_{n,2n}+\frac{\ld}{4}e_{n+1,2n}
\nn\\ &+
(-1)^n\big(e_{2n,n+1}-\frac{1}{2}e_{2n,n}-\frac{\ld}{2}e_{1,n+1}+\frac{\ld}{4}e_{1,n}\big)\Big)
\label{Gm}
\end{align}
with~$\kappa=1$ when~$n$ is even and~$\sqrt{-1}$ when~$n$ is odd.
Here $\Ld^j$ and $\Gm^j$ are define to be the $j$-th power of $\Ld$
and $\Gm$ respectively for $j>0$, while for $j<0$
\begin{equation}\label{LdGm}
 \Ld^j=(\ld^{-1}\Ld^{2n-3})^{-j},\quad
 \Gm^j=(\ld^{-1}\Gm)^{-j}.
\end{equation}

We now rewrite the Drinfeld-Sokolov hierarchy of $D_n$ type \eqref{dnm} into the form
\begin{equation}\label{DS-pn}
\frac{\pd \msL}{\pd t_k}=[\varphi(-\Ld^k)^+, \msL], \quad
\frac{\pd \msL}{\pd \hat{t}_k}=[\varphi(\Gamma^k)^+, \msL],  \quad k\in\Zop.
\end{equation}
We call the flows $\frac{\p}{\p t_k}$ and $\frac{\p}{\p \hat{t}_k}$ {\em the
positive and the negative flows of the Drinfeld-Sokolov hierarchy of $D_n$ type } respectively. We will show that these flows coincide with the positive and negative flows \eqref{dn1} and \eqref{dn2} defined by the pseudo-differential operator $L$.

It is shown in \cite{DS} that in the orbit of gauge transformations
of $\msL$, one can find a canonical representative
$\msL^{\mathrm{can}}$ of the form
\[\msL^{\mathrm{can}}=D+\Ld+q^{\mathrm{can}},\]
where $q^{\mathrm{can}}$ reads
\begin{align}\label{qcan}
q^{\mathrm{can}}=&\sum_{j=1}^{[\frac{n-1}{2}]}\Big(  q_j (e_{1,2j}+e_{2n+1-2j,
2n})+ q_{n-j}(e_{1,2n+1-2j}+e_{2j, 2n}) \Big) + \hat{q}
\end{align}
with
\begin{equation*}\label{}
\hat{q}=\left\{ \begin{array}{ll} \frac{1}{2}( q_{n/2}+\rho)
(e_{1,n}+e_{n+1, 2n})+
( q_{n/2}-\rho) (e_{1,n+1}+e_{n, 2n}), & n \hbox{ even}, \\
- \sqrt{-1}\rho (\frac{1}{2}e_{1,n}- e_{1,n+1}+ e_{n,2n} -
\frac{1}{2} e_{n+1, 2n}), & n \hbox{ odd}.
\end{array}\right.
\end{equation*}
The coefficients $q_1, \dots, q_{n-1}$ and $\rho$ are gauge
invariant differential polynomials of $q$ that appears in
\eqref{msL}. They serve as coordinates of the orbit space of gauge
transformations, and we will use them as unknown functions of the
Drinfeld-Sokolov hierarchy.

Let $\mA$ be the algebra of differential polynomials of $q_1, \dots,
q_{n-1}$ and $\rho$, denote $\mA^-=\mA((\lambda^{-1}))$, we
introduce a free $\mA^-$-module
\[
V=\left(\mA^-\right)^{2n}=\left\{\sum_{i <\infty} \al_i \ld^i\mid
\al_i\in \mA^{2n} \right\}.
\]
Let us fix a basis $\{\hat{\psi}_{2n}, \psi_{2n-1},
\dots,\psi_{1}\}$ of $V$, where
$\hat{\psi}_{2n}=\frac{\ld}{2}\psi_1+\psi_{2n}$, and $\psi_i$ is the
column vector whose $i$-th entry is $1$ and others are zero.

In the notions of Lemma \ref{lemma-Dt}, we let $\mR=\mD^-$ and
denote by $\mR_+$ the subalgebra of $\mR$ consisting of differential
operators. We define an $\mR_+$-module structure on $V$ by
\begin{equation}\label{LV}
 D\cdot\al=\msL^{\mathrm{can}}\al,~~ \al\in V.
\end{equation}
Note that $\left.\msL^{\mathrm{can}}\right|_{\lambda=0}\in\mR_+^{2n\times 2n}$, let
\[
R=\left(\left.\msL^{\mathrm{can}}\right|_{\lambda=0}\right)^T
=-\diag(D,D,\dots,D)+\Ld|_{\lambda=0}+\left(q^{\mathrm{can}}\right)^T,
\]
then it is straightforward to verify  that
\begin{equation}\label{l0}
R\cdot(\hat{\psi}_{2n}, \psi_{2n-1},\ldots, \psi_1)^t= (-\ld\psi_2,
0,\ldots, 0)^t=(-\ld D\cdot\psi_1, 0,\ldots, 0)^t.
\end{equation}

Denote $\mL=-\Dt(R)$, where $\Dt$ is the operation defined in Lemma
\ref{lemma-Dt}, then $\mL^*=-\mL$ by using \eqref{g} and the third
part of Lemma~\ref{lemma-Dt}. It is easy to see that $\mL$ has the
form \eqref{mL}. This observation gives a Miura-type transformation
between $u^1, \dots, u^n$ and $q_1, \dots, q_{n-1}, \rho$, so the
algebra $\mA$ defined above coincides with the one that is given in the last section.
Moreover, the second part of Lemma \ref{lemma-Dt} implies that $\mL$
is invariant w.r.t. the gauge transformations \eqref{gauge}, thus
the Drinfeld-Sokolov hierarchy can be represented by the operator
$\mL$, or equivalently by $L=D^{-1}\mL$.

Note that the operator $\mL\notin \mR_+$, since $V$ is only an
$\mR_+$-module $\mL$ cannot act on $V$, and the first part of
Lemma \ref{lemma-Dt} cannot be applied directly. To resolve this problem,
Drinfeld and Sokolov decomposed $V$ into two subspaces such that
$\mD^-$ can act on one of them, then the first part of Lemma
\ref{lemma-Dt} can be applied. In this way,
the positive flows of the Drinfeld-Sokolov hierarchy \eqref{DS-pn}
are represented in the form \eqref{dn1}
as the positive flows given by the pseudo-differential operataor $L$ of the form \eqref{eq-LL}.

In the matrix realization of $\mfg$, the elements $\Ld$ and $\Gm$
are $2n\times 2n$ matrices with entries in $\C[\ld]$, so they can
act on the space $V$. One can verify that the following
decomposition holds true
\[V=V_1\oplus V_2, \quad V_1= \im\,\Ld=\kn\,\Gm, \quad V_2= \kn \Ld=\im\,\Gm.\]
Denote $T=e^U$, where $U$ is the matrix appeared in Proposition
\ref{prop-drg} with $\msL=\msL^{\mathrm{can}}$, then we also have
\begin{equation}\label{v}
V=V_1'\oplus V_2', \quad V_1'=TV_1, \quad V_2'=TV_2.
\end{equation}

Since the operator $\ld^{-1}\Ld^{2n-2}$ is the identity operator when
restricted to  $V_1$, let
$\sP=\varphi(\ld^{-1}\Ld^{2n-2})$ with $\varphi$ being defined in
\eqref{phi}, then $\sP$ is the projection from $V$ to $V_1'$. We
denote the projection of $\al\in V$ in $V_1'$ by $\al'=\sP \al$, and
define the action
\begin{align*}
& D^{-1}\cdot\al'=\left(\msL^{\mathrm{can}}\right)^{-1}\al'= T \big(\Ld-(\Ld-\mathscr{L}_0)\big)^{-1} T^{-1}\al'\\
=&  T \big(\Ld^{-1}+\Ld^{-1}(\Ld-\mathscr{L}_0)\Ld^{-1}
+(\Ld^{-1}(\Ld-\mathscr{L}_0))^2\Ld^{-1}+\cdots\big) T ^{-1}\al'.
\end{align*}
Here the operator $\msL_0$ defined in \eqref{L0g} now reads
\begin{equation}\label{L0}
\msL_0=e^{-U}\msL^\mathrm{can} e^U= D+\Ld+\sum_{k\in\Zop}f_k
\Ld^{-k}+\sum_{k\in\Zop}g_k\Gm^{-k}
\end{equation}
with $f_k, g_k\in\mA$ and the negative powers $\Ld$, $\Gm$ defined
in  \eqref{LdGm}. Note that $\im\,\Ld^{-1}\subset \im\,\Ld$,
$\im\,\Gm^{-1}\subset\ker\,\Ld$, then $D^{-1}\cdot\alpha'\in V_1'$,
so $V_1'$ becomes an $\mR$-module.

It follows from $[\msL_0, \Ld]=0$ that $[\sP,
\msL^{\mathrm{can}}]=0$, then by acting $\sP$ on both sides of
\eqref{l0} one has
\begin{equation*}\label{l0p}
R\cdot(\hat{\psi}_{2n}', \psi_{2n-1}',\ldots, \psi_1')^t= (-\ld
D\cdot\psi_1', 0,\ldots, 0)^t.
\end{equation*}
Now the first part of Lemma~\ref{lemma-Dt} can be employed to prove
the following lemma.
\begin{lemma}[\cite{DS}] \label{thm-Ppsi1}
Let $\mL=-\Dt(R)$, $L=D^{-1}\mL$, then $L$ takes the form
\eqref{eq-LL}. Define $P=L^{\frac1{2n-2}}\in \mD^-$ as in
Lemma~\ref{lem-3-1}, then for any $i\in\Z$ the following equalities
hold true
\begin{align}
&\varphi(\Ld^{i})\psi_1'=P^{i}\cdot\psi_1', \\
& \big(\varphi(\Ld^{2i+1})^+\psi_1\big)'=(P^{2i+1})_+\cdot\psi_1'. \label{md1}
\end{align}
\end{lemma}

By using the second equality, one can represent the
positive flows $\frac{\p}{\p t_k}$ of the Drifeld-Sokolov hierarchy \eqref{DS-pn}
in the form \eqref{dn1}.
We are to explain in the next subsection that the negative flows of \eqref{DS-pn}
can be represented as \eqref{dn2}.

The first equality of the above lemma gives the following result.
\begin{prop}[\cite{DS}]\label{hd1}
Let $f_k$ be the coefficients that appear in \eqref{L0}, then
$f_k+\frac{1}{k}\res\, P^{k}\in D(\mA)$ for all $k\in\Zop$.
\end{prop}
From Theorem \ref{th-ds} and \eqref{HkHhk} we know that this proposition related the densities of the Hamiltonians of the positive flows of the Drinfeld-Sokolov hierarchy with that of the positive flow \eqref{dn1} defined in the last section.

\subsection{Negative flows of the Drinfeld-Sokolov hierarchy of $D_n$ type}

In the last subsection, the pseudo-differential operator representation
for the positive flows of the Drinfeld-Sokolov hierarchy of $D_n$
type is obtained by introducing a $\mD^-$-module structure on the
space $V_1'$ and using Lemma \ref{lemma-Dt} as was done in
\cite{DS}. In order to obtain a similar representation for the
negative flows, we try to assign a $\mD^+$-module structure to
$V_2'$. However, it seems that there is no such a structure on
$V_2'$, so we first extend the space $V_2'$ to a larger one $V_2''$
which admits a $\mD^+$-module structure, then we employ Lemma
\ref{lemma-Dt} and obtain the pseudo-differential operator
representation for the negative flows of the Drinfeld-Sokolov
hierarchy of $D_n$ type.

Recall that $V_2$ as an $\mA^-$-module is spanned by the following two vectors:
\begin{equation}\label{zh-12-8-1}
\hat{\psi}_1=\frac{1}{2}\psi_1-\frac{1}{\ld}\psi_{2n}, \quad
\hat{\psi}_2=\Gm\hat{\psi}_1=\ka\left(\psi_n-\frac{1}{2}\psi_{n+1}\right).
\end{equation}
The action of $\Gm$ restricted to $V_2$ satisfies $\Gm^2=\ld$, so we
introduce $\Gm^{-1}=\ld^{-1} \Gm$, see \eqref{LdGm}. It is easy to
see that every vector $\al\in V_2$ can be uniquely expressed in the
form
\begin{equation}
\al=\sum_{i\leq m}a_i\Gm^i\hat\psi_1, \quad  a_i\in\mA, ~~m\in\Z. \label{eq-alpha}
\end{equation}
This observation shows that the space $V_2$ is in fact a rank-one free module of the following algebra
\[\mD^-(\mA, \Gm)=\left\{\sum_{i<\infty}a_i\Gm^i\mid
a_i\in\mA\right\}.
\]
This is the algebra of ``pseudo-differential operators of the first
type'' (see Sec. 2.1) over the algebra $\mA$ with the derivation
``$D$'' being the following trivial map
\[\Gm:\mA\to\mA,\quad f \mapsto 0,\]
which surely gives a derivation of degree one over $\mA$.

By regarding another trivial map
\[\Gm^{-1}:\mA\to\mA,\quad f \mapsto 0,\]
as a derivation of degree one, one can also define the algebra of
``pseudo-differential operators of the second type'' with respect to
the algebra $\mA$ and the derivation $\Gm^{-1}$ as
\[\mD^+(\mA, \Gm^{-1})=\left\{\sum_{j\geq0}\sum_{i\leq m+j}a_{i,j}\Gm^i\mid
a_{i,j}\in\mA_j, ~m\in\Z\right\}.
\]
We denote by $\hat{V}_2$ the rank-one free module of the algebra
$\mD^+(\mA, \Gm^{-1})$ with generator $\hat{\psi}_1$, which has a
linear topology induced from that of $\mD^+(\mA, \Gm^{-1})$. It is
easy to see that the algebra $\mD^-(\mA, \Gm)$ is a subalgebra of
$\mD^+(\mA, \Gm^{-1})$ (see Lemma \ref{lem-supp}), hence $V_2$ is a
subspace of $\hat{V}_2$.

To define the space $V_2''$, we need to extend the space $V$ to
certain space $\hat{V}$ that involves $\hat{V}_2$ as a subspace.
Since the space $V$ is defined to be $\left(\mA^-\right)^{2n}$, in
which the algebra $\mA^-=\mA((\ld^{-1}))$ can also be defined as
$\mD^-(\mA, \ld)$ with $\ld$ being the trivial derivation, we
similarly extend the space $V$ to
\[\hat{V}=\hat{\mA}^{2n},\quad \hat{\mA}=\mD^+(\mA, \ld^{-1}).\]
The space $\hat{V}$ has a linear topology induced from that of
$\hat{\mA}$. It is easy to see that the linear transformations $\Ld,
\Gm, T=e^U: V \to V$ can be extended naturally to $\hat{V}$. Then
the expression
\begin{equation}\label{alV2}
\al=\sum_{j\geq0}\sum_{i\leq m+j}a_{i,j}\Gm^i\hat\psi_1\in\hat{V}_2
\end{equation}
is also convergent in $\hat{V}$ according to its topology, hence the
space $\hat{V}_2$ is indeed a subspace of $\hat{V}$.

Now let us introduce another subspace of $\hat{V}$:
\[V_2''=T\,\hat{V}_2\subset \hat{V},\]
then $V_2'$ is a subspace of $V_2''$. As in the previous subsection
we define a map
\[\sQ: V\to V_2', \quad \sQ=\varphi(\ld^{-1}\Gm^2) \]
with $\varphi$ defined in \eqref{phi}. Then we have the following
commutative diagram
\[
\xymatrix{ V  \ar[rr]^{\ld^{-1}\Gm^2}  \ar[d]^T_\cong & & V_2\, \ar@{^(->}[rr]^i \ar[d]^T_\cong && \hat{V}_2 \ar[d]^T_\cong \\
V \ar[rr]^{\sQ} && V_2'\, \ar@{^(->}[rr]^i & & V_2''  }
\]
We also denote the composition of $\sQ$ and the inclusion
$V_2'\hookrightarrow V_2''$ by $\sQ$,  and write $\al''=\sQ\al$ for
any vector $\al\in V$.

\begin{lemma}\label{thm-alV22}
The space $\hat{V}_2$ is a free $\mD^+(\mA,\Gm^{-1})$-module with
generator $T^{-1}\psi_1''$.
\end{lemma}
\begin{prf}
To see that $T^{-1}\psi_1''$ is another generator besides
$\hat{\psi}_1$, we only need to show that these two vectors are
related by the action of a unit of the algebra
$\mD^+(\mA,\Gm^{-1})$.

Recall $T=e^U$, in which according to the present matrix realization
the element $U$ given in Proposition \ref{prop-drg} has the form
$U_0+O(\ld^{-1})$ with $U_0$ being a strictly upper triangular
matrix, and that the vector $\hat\psi_1$ defined in
\eqref{zh-12-8-1} can be represented as
\[\hat\psi_1=\ld^{-1} \Gm^2 \psi_1,\]
so we have
\[
\psi_1''=\sQ\psi_1=T \ld^{-1} \Gm^2 T^{-1} \psi_1=T (\hat{\psi}_1+O(\ld^{-1}))\in V_2'.
\]
By using the general form  \eqref{eq-alpha} of elements of $V_2$ and
the identity $\Gm^{2j+1}|_{V_2}=\ld^j \Gm$, one can represent
$T^{-1}\psi_1''\in V_2$ in the following form:
\begin{equation}\label{Tpsi}
T^{-1}\psi_1''=\left(1+\sum_{i<0}b_i\Gm^i\right)\hat{\psi}_1, \quad
b_i\in\mA.
\end{equation}
Obviously the element $1+\sum_{i<0}b_i\Gm^i\in\mD^+(\mA,\Gm^{-1})$
is invertible. The lemma is proved.
\end{prf}

Aiming at a $\mD^+$-module structure on the space $V_2''$ such that
the action of $D$ coincides with \eqref{LV} when restricted to the
subspace $V_2'$, we need to define the action of
$(\msL^{\mathrm{can}})^i$ $(i\in\Z)$ on the space $V_2''$. Note that
the operator $\msL_0:V \to V$ given in \eqref{L0} can be extended to
$\hat{V}$, we denote its restriction on the space $\hat{V}_2$ by
$\hat{\msL}_0$, which reads
\[\hat{\msL}_0=\msL_0|_{\hat{V}_2}=D+\sum_{k\in\Zop}g_k\Gm^{-k}.
\]
Here $g_1\in\mA$ is invertible as indicated in \cite{DS}, so the
operator $\hat{\msL}_0$ is invertible on $\hat{V}_2$, and its
inverse is given by
\begin{align*}\label{}
\hat{\msL}_0^{-1}=&\big(g_1\Gm^{-1}(1+g_1^{-1}\Gm\,D+M)\big)^{-1}\nn\\
=&\big(1-(g_1^{-1}\Gm\,D+M)+(g_1^{-1}\Gm\,D+M)^2-\cdots\big)g_1^{-1}\Gm,
\label{L0inv}
\end{align*}
where $M=g_1^{-1}\sum_{j\ge1}g_{2j+1}\,\Gm^{-2j}$. One can expand
the right hand side and obtain
\begin{equation}\label{L0inv}
\hat{\msL}_0^{-1}=\sum_{s\geq0}\sum_{r\leq s}A_{rs}\,\Gm^{r+1},
\quad A_{rs}=\sum_{j=0}^s c_{rsj}D^j,~~ c_{rsj}\in\mA_{s-j},
\end{equation}
in which $A_{00}=c_{000}=g_{10}^{-1}$ with $g_{10}$ being the
projection of $g_1$ onto $\mA_0$. Note that $g_{10}/\rho$ is a
positive constant, where $\rho$ appears in the definition
\eqref{qcan} of $\msL^{\mathrm{can}}$, and we have normalized $\Gm$
such that this constant is $1$. Since $A_{rs}$ are differential
operators of degree $s$, i.e., $A_{rs}(\mA_d)\subset\mA_{d+s}$, then
by using the expressions \eqref{L0inv} and \eqref{alV2} one can
verify that the action of $\hat{\msL}_0^{-1}$ on $\hat{V}_2$ is well
defined. Also note that the image $\hat{\msL}_0^{-1}(V_2)$ is not
contained in $V_2$ though $\hat{\msL}_0(V_2)\subset V_2$, which is
why we extend $V_2$ to $\hat{V}_2$.

To go forward, we need to present another expression for vectors in
$\hat{V}_2$.
\begin{lemma}\label{thm-V2}
Every vector $\al\in \hat{V}_2$ can be uniquely expressed in the
form
\begin{equation}\label{V2L0}
\al=\sum_{j\geq0}\sum_{i\leq
m+j}b_{i,j}\hat{\msL}_0^{-i}\,T^{-1}\psi_1'', \quad b_{i,j}\in\mA_j,
~~m\in\Z.
\end{equation}
\end{lemma}
\begin{prf}
According to Lemma \ref{thm-alV22}, we suppose $\al \in \hat{V}_2$
has the form
\[\al=\sum_{j\ge k}\sum_{i\le m+j}a_{i,j}\Gm^i\,T^{-1}\psi_1''+\cdots, \quad a_{i,j}\in\mA_j,\]
where $\cdots$ stands for the terms of the form \eqref{V2L0}. Let us
proceed to prove the lemma by induction on the lower bound $k$ of
the index $j$.

First, we have
\begin{align}\label{aldec}
\al=&\sum_{i\le m+k}a_{i,k}\Gm^i\,T^{-1}\psi_1''+\sum_{j\ge
k+1}\sum_{i\le m+j}
a_{i,j}\Gm^iT^{-1}\psi_1''+\cdots \nn\\
=&a_{m+k,k}\Gm^{m+k}\,T^{-1}\psi_1''+\sum_{i\le m-1+k}a_{i,k}\Gm^i\,T^{-1}\psi_1''\nn\\
&\quad+\sum_{j\ge k+1}\sum_{i\le m+j}
a_{i,j}\Gm^i\,T^{-1}\psi_1''+\cdots.
\end{align}
From the expansion \eqref{L0inv} it follows that
\begin{equation}
\hat{\msL}_0^{-l}=\sum_{s\geq0}\sum_{r\leq s}A^{(l)}_{rs}\,\Gm^{r+l},
\quad A^{(l)}_{rs}\in(\mD^-)_+, ~~\deg A^{(l)}_{rs}=s, \label{l0hat}
\end{equation}
where $A^{(l)}_{00}=g_{10}^{-l}$, hence by using \eqref{Tpsi} we
have
\begin{align}
& \hat{\msL}_0^{-l}T^{-1}\,\psi_1''-g^{-l}_{10}\Gm^l\,T^{-1}\psi_1'' \nn\\
=& \left(\sum_{r\le-1}A^{(l)}_{r0}\Gm^{r+l}+\sum_{s\ge1}\sum_{r\le
s}A^{(l)}_{rs}\Gm^{r+l}\right)T^{-1}\psi_1'' \nn\\
=& \left(\sum_{r\le-1}A^{(l)}_{r0}\Gm^{r+l}+\sum_{s\ge1}\sum_{r\le
s}A^{(l)}_{rs}\Gm^{r+l}\right)\left(1+\sum_{i<0}b_i\Gm^i\right)\hat{\psi}_1
\nn\\
=& \left(\sum_{r\le-1}c_{r,0}\Gm^{r+l}+\sum_{s\ge1}\sum_{r\le
s}c_{r,s}\Gm^{r+l}\right)\hat{\psi}_1 \nn\\
=&
\left(\sum_{r\le-1}\tilde{c}_{r,0}\Gm^{r+l}+\sum_{s\ge1}\sum_{r\le
s}\tilde{c}_{r,s}\Gm^{r+l}\right)T^{-1}\psi_1'',\label{L0Gm}
\end{align}
where $c_{r,s}, \tilde{c}_{r,s}\in\mA_s$. The above computation
represents the action of the operator $\hat{\msL}_0^{-l}$
\eqref{l0hat} on certain vector in $\hat{V}_2$ by an element in
$\mD^+(\mA, \Gm^{-1})$. By using equation \eqref{L0Gm}, we can
eliminate the term $a_{m+k,k}\Gm^{m+k}\,T^{-1}\psi_1''$ in
\eqref{aldec} and arrive at
\[
\al=\sum_{i\le m-1+k}\tilde{a}_{i,k}\Gm^i\,T^{-1}\psi_1''+\sum_{j\ge
k+1}\sum_{i\le m+j}
\tilde{a}_{i,j}\Gm^i\,T^{-1}\psi_1''+\cdots,\quad
\tilde{a}_{i,j}\in\mA_j.
\]

Then by induction on the upper bound of the index $i$ appearing in
the first summation we have
\[
\al=\sum_{j\ge k+1}\sum_{i\le m+j}
\tilde{\tilde{a}}_{i,j}\Gm^i\,T^{-1}\psi_1''+\cdots,
\]
which shows that the lower bound of the index $j$ has increased by
one. The lemma is proved.
\end{prf}

Now we are ready to introduce a $\mD^+$-module structure on the
space $V_2''$ by defining the action
\begin{equation}\label{Dal}
 D^{i}\cdot\al''=\varphi(\hat{\mathscr{L}}_0^{i})\al'',
\quad\al''\in V_2'',~~i\in\Z,
\end{equation}
which extends the action \eqref{LV} on  $V_2'$ to an action on
$V_2''$. Then Lemma \ref{thm-V2} is equivalent to the following
theorem.
\begin{thm}\label{thm-V22}
The  $\mD^+$-module $V_2''$ is a free module with generator $\psi_1''$.
\end{thm}

Let us apply Lemma \ref{lemma-Dt} to the algebra $\mR=\mD^+$ and the
module $V_2''$. By acting the projection operator $\sQ$ to both
sides of \eqref{l0}, we have
\begin{equation*}\label{l02}
R\cdot(\hat{\psi}_{2n}'', \psi_{2n-1}'',\ldots, \psi_1'')^t=(-\ld
D\cdot\psi_1'', 0,\ldots, 0)^t,
\end{equation*}
hence $L\cdot\psi_1''=\ld\,\psi_1''$, where $L=-D^{-1}\Dt(R)$ as
given before. According to Lemma \ref{lemma-lq} we introduce a
pseudo-differential operator $Q\in \mD^+$ such that $L=Q^2$, and
consider the action of $Q^i$ on $V_2''$ for any integer $i$.
\begin{lemma} \label{lem-GmQ}
For any integer $i$ the following equality holds true:
\begin{equation}
\varphi({\Gm}^{i})\psi_1''=Q^{i}\cdot\psi_1''.
\end{equation}
\end{lemma}
\begin{prf} We only need to prove the case $i=1$.
Since $V_2''$ is a free $\mD^+$-module, there exists an element
$A\in\mD^+$ such that $\varphi(\Gm)\psi_1''=A\cdot\psi_1''$. Note
that $[\varphi(\Gm),\mathscr{L}^{\mathrm{can}}]=0$, so the action of
$\varphi(\Gm)$ on $V_2''$ commutes with $D\in\mD^+$, hence
\[A^2\cdot\psi_1''=\varphi(\Gm^{2})\psi_1''= \ld\psi_1''=L\cdot\psi_1''.
\]
By using the freeness of $V_2''$, we have $A^2=L=Q^2$. It follows that $A=\pm Q$.

To show $A=Q$, we only need to compare their leading terms. Equation
\eqref{L0Gm} leads to
\[
\varphi(\Gm)\psi_1''=\varphi(g_{10}\hat{\msL}_0^{-1}+\cdots)\psi_1''
=(g_{10} D^{-1}+\cdots)\cdot\psi_1'',
\]
which implies that the leading term of $\res\, A$ is $g_{10}$. On
the other hand $g_{10}$ takes the same sign with $\rho=\res\, Q$,
thus $A=Q$. The lemma is proved.
\end{prf}

By using Lemmas \ref{thm-res} and \ref{lem-GmQ}, one can prove the
following proposition. The argument is almost the  same with the one
for Proposition \ref{hd1} in \cite{DS}, so we omit the details here.
\begin{prop}\label{hd2}
Let $g_k$ be the coefficients that appear in \eqref{L0}, then
$g_k-\frac{1}{k}\res\, Q^{k}\in D(\mA)$ for all $k\in\Zop$.
\end{prop}
This proposition connects the Hamiltonians of the negative flows of the Drinfeld-Sokolov hierarchy of $D_n$ type to those \eqref{HkHhk} corresponding to the
negative flows \eqref{dn2}.

Now we arrive at the main result of the present section.
\begin{thm}\label{thm-main}
The flows \eqref{DS-pn} of the Drinfeld-Sokolov hierarchy of $D_n$ type coincide with the flows of the integrable
hierarchy \eqref{dn1}, \eqref{dn2}.
\end{thm}
\begin{prf}
It is shown in \cite{DS} that the Drinfeld-Sokolov hierarchy of
$D_n$ type has a bihamiltonian structure given by the two Poisson
brackets \eqref{poi1}, \eqref{poi2}. For the flow \eqref{dnm}
corresponding to the element $\ld_j$, the Hamiltonian with respect
to the second Poisson bracket is given by
\[
\mathcal{H}_j=\int (H\mid\ld_j)\od x, \quad j\in E_+,
\]
where $H$ is given in \eqref{L0g} and $(\cdot\mid\cdot)$ is the
trace form defined by
\[(G\mid H)=\res_{\ld}\left(\ld^{-1}\tr(G\,H)\right).\]
We choose a basis \eqref{base-s} of the Heisenberg subalgebra $\mathfrak{s}$. as
\[
\ld_k=-\Ld^k, \quad \ld_{k(n-1)'}=\Gm^k, \quad k\in 2\Z+1.
\]
Note that
\[
(\Ld^k\mid\Ld^l)=(2n-2)\dt_{k,-l},\quad (\Ld^k\mid\Gm^l)=0, \quad
(\Gm^k\mid\Gm^l)=2\,\dt_{k,-l},
\]
where $k$, $l$ run over all odd integers, hence by using \eqref{L0}
we have
\[
\mathcal{H}_{k}=-(2n-2)\int\!\! f_k\,\od x,\quad
\mathcal{H}_{k(n-1)'}=2\int\!\! g_k\,\od x,\quad k\in\Zop.
\]
They are the Hamiltonians for the positive and negative flows of the Drinfeld-Sokolov hierarchy \eqref{DS-pn} w.r.t. the second Poisson bracket \eqref{poi2}.

 According to Propositions \ref{hd1}, \ref{hd2} and
Theorem~\ref{thm-bh}, these Hamiltonians satisfy
\[
\mathcal{H}_k=H_{k}, \quad \mathcal{H}_{k(n-1)'}=\hat{H}_{k},\quad
k\in\Zop,
\]
where $H_k$, $\hat{H}_k$ are the Hamiltonians of the integrable
hierarchy \eqref{dn1}, \eqref{dn2} with respect to the second
Poisson bracket \eqref{poi2}. So the Drinfeld-Sokolov hierarchy of
$D_n$ type \eqref{DS-pn} and the integrable hierarchy \eqref{dn1},\eqref{dn2}
coincide. The theorem is proved.
\end{prf}

\section{The two-component BKP hierarchy and its reductions}
In this section we represent the two-component BKP hierarchy that is introduced in
\cite{DJKM-KPtype} via pseudo-differential operators, and show that
the hierarchy \eqref{dn1}, \eqref{dn2} is just a reduction, which was considered in \cite{DJKM-reduce}, of the
two-component BKP hierarchy.

\subsection{The two-component BKP hierarchy}
Let $\tilde{M}$ be an infinite-dimensional manifold with local
coordinates
\[(a_1, a_3, a_5, \dots, b_1, b_3, b_5, \dots),\]
and $\tilde{\mA}$ be the algebra of differential polynomials on
$\tilde{M}$:
\[\tilde{\mA}=C^{\infty}(\tilde{M})[[a_{i}^{s}, b_{i}^{s}\mid i\in\Zop, s\ge1]].\]
As in Section 3, we assign a gradation on $\tilde{\mA}$ such that
$\tilde{\mA}$ is topologically complete.
Define a derivation $D$ by
\[D=\sum_{s\ge0}\sum_{i\in \Zop} \left(a_i^{s+1}\frac{\p}{\p a_i^s}+b_i^{s+1}\frac{\p}{\p b_i^s}\right),\]
then the algebras
$\tilde{\mD}^{\pm}=\mD^\pm(\tilde{\mA},D)$ of pseudo-differential
operators can be constructed as we did in Section 2.1.

Introduce two pseudo-differential operators
\begin{align}
\Phi=&1+\sum_{i\ge 1}a_i  D^{-i} \in \tilde{\mD}^-,\\
\Psi=&1+\sum_{i\ge 1}b_i  D^{i} \in \tilde{\mD}^+,
\end{align}
where $a_2, a_4, a_6, \dots, b_2, b_4, b_6, \dots\in \tilde{\mA}$ are determined by the following condtions
\begin{equation}
\Phi^*=D\Phi^{-1}D^{-1},\quad \Psi^*=D\Psi^{-1}D^{-1}.
\label{phipsi}
\end{equation}

Now let us define a pair of operators
\begin{equation*}
P=\Phi D \Phi^{-1} \in \tilde{\mD}^-,\quad Q=\Psi D^{-1} \Psi^{-1}
\in \tilde{\mD}^+.
\end{equation*}
\begin{lemma}\label{lempq}
The operators $P, Q$ have the following expressions (c.f.
\eqref{eq-P}, \eqref{q}):
\begin{equation*}
P=D+\sum_{i\ge1}u_i D^{-i}, \quad Q=D^{-1}\rho+\sum_{i\ge1} v_i D^i,
\end{equation*}
where $\rho=(\Psi^{-1})^*(1)$. They satisfy
\begin{equation}\label{PQstar}
P^*=-D P D^{-1}, \quad Q^*=-D Q D^{-1},
\end{equation}
and that for any $k\in\Zop$
\begin{equation}
(P^k)_+(1)=0,\quad (Q^k)_+(1)=0. \label{PkQk}
\end{equation}
\end{lemma}
\begin{prf}
The expression of $P$ is obvious. To show that of $Q$, we consider
its negative part:
\begin{align*}
Q_-=& \left(\Psi D^{-1} \Psi^{-1}\right)_-=\left(D^{-1} \Psi^{-1}\right)_-=\left(\left(D^{-1} \Psi^{-1}\right)^*\right)_-^* \\
=&-\left((\Psi^{-1})^*D^{-1}\right)_-^*=-\left((\Psi^{-1})^*(1)D^{-1}\right)^*=D^{-1}\rho.
\end{align*}
The symmetry property \eqref{PQstar} is obvious, which implies
\eqref{PkQk}. The lemma is proved.
\end{prf}

We define the following evolutionary equations: 
\begin{align}
&\frac{\pd \Phi}{\pd t_k}=- (P^k)_-\Phi, \quad \frac{\pd \Psi}{\pd t_k}=\bigl((P^k)_+ -\dt_{k1} Q^{-1}\bigr)\Psi, \label{ppt1}\\
&\frac{\pd \Phi}{\pd \hat{t}_k}=- (Q^k)_-\Phi, \quad \frac{\pd \Psi}{\pd \hat{t}_k}=(Q^k)_+\Psi, \label{ppt2}
\end{align}
where $k\in\Zop$. According to \eqref{phipsi} and \eqref{PkQk}, it
is easy to see that these flows are well defined, and they yield the
Lax equations of the form \eqref{Pt}, \eqref{Qt}. By a
straightforward calculation one can verify the commutativity of
these flows, hence they form an integrable  hierarchy indeed. We will show that
 this hierarchy possesses tau functions, and
that these tau functions satisfy the same bilinear equations of the
two-component BKP hierarchy defined in \cite{DJKM-KPtype}.

First, let us introduce two wave functions
\begin{align}\label{wavef}
w=w(\mathbf{t}, \hat{\mathbf{t}}; z)=\Phi e^{\xi(\mathbf{t};z)},
\quad \hat{w}=\hat{w}(\mathbf{t}, \hat{\mathbf{t}}; z)=\Psi
e^{x z+\xi(\hat{\mathbf{t}};-z^{-1})},
\end{align}
where $x=t_1$, the function $\xi$ is defined by
\begin{equation}\label{xi}
\xi(\mathbf{t}; z)=\sum_{k\in\Zop} t_k z^k,
\end{equation}
and for any $i\in\Z$ the action of $D^i$ on $e^{x z}$ is set to be
$D^i e^{x z}=z^i e^{x z}$.

It is easy to see that $P\,w=z w, \quad Q\, \hat{w} = z^{-1}
\hat{w}$, and that the flows \eqref{ppt1}, \eqref{ppt2} are
equivalent to the following equations
\begin{align}
&\frac{\pd w}{\pd t_k}=(P^k)_+ w, \quad \frac{\pd \hat{w}}{\pd t_k}=(P^k)_+ \hat{w}, \\
&\frac{\pd w}{\pd \hat{t}_k}=- (Q^k)_- w, \quad\frac{\pd \hat{w}}{\pd \hat{t}_k}=-(Q^k)_- \hat{w}.
\end{align}
Here $(Q^k)_- w$ is understood as $\left((Q^k)_-
\Phi\right)e^{\xi(\mathbf{t};z)}$, and $(Q^k)_- \hat{w}$ is defined  similarly.
The following theorem can be proved as it was done for the KP hierarchy given in
\cite{DKJM-KPBKP, Dickey}.
\begin{thm}\label{thm-bl}
The hierarchy \eqref{ppt1}, \eqref{ppt2} is equivalent to the following bilinear equation
\begin{equation}\label{bl2BKP}
\res_z z^{-1} w(\mathbf{t},\hat{\mathbf{t}}; z)w(\mathbf{t}',
\hat{\mathbf{t}}'; -z) =\res_z z^{-1}\hat{w}(\mathbf{t},
\hat{\mathbf{t}}; z) \hat{w}(\mathbf{t}',\hat{\mathbf{t}}'; -z).
\end{equation}
\end{thm}
Here and below the residue of a Laurent series is defined as
$\res_z\sum_i f_i z^i=f_{-1}$.

Let $\omega$ be the following $1$-form
\begin{equation}
\omega=\sum_{k\in\Zop} \left(\res\,P^k\,\od
t_k+\res\,Q^k\,\od\hat{t}_k\right). \label{tau-1}
\end{equation}
By using the equations \eqref{ppt1} and \eqref{ppt2}, one can show
that $\omega$ is closed, so given any solution of the hierarchy
\eqref{ppt1}, \eqref{ppt2} there exists a function
$\tau(\mathbf{t},\hat{\mathbf{t}})$ such that
\begin{equation}
\omega=\od\left(2\,\p_x\,\log\tau\right). \label{tau-2}
\end{equation}
Moreover, one can fix a
tau function such that the wave functions can be written as
\begin{align}
&w(\mathbf{t},\hat{\mathbf{t}};z)=\frac{\tau(\ldots, t_k-\frac{2}{k z^k}, \ldots, \hat{\mathbf{t}})}{\tau(\mathbf{t},\hat{\mathbf{t}})}
e^{\xi(\mathbf{t};z)}, \label{wtau1}\\
&\hat{w}(\mathbf{t},\hat{\mathbf{t}};z)=\frac{\tau(\mathbf{t},\ldots, \hat{t}_k+ \frac{2z^k}{k},\ldots)}{\tau(\mathbf{t},\hat{\mathbf{t}})}
e^{\xi(\hat{\mathbf{t}};-z^{-1})}. \label{wtau2}
\end{align}

Introduce a vertex operator $X$ as
\[X(\mathbf{t};z)=\exp\left(\sum_{k\in\Zop} t_k z^k\right)\exp\left(-\sum_{k\in\Zop}\frac{2}{k z^k}\frac{\p}{\p t_k}\right),\]
then the bilinear equation \eqref{bl2BKP} reads
\begin{align*}
&\res_z z^{-1}X(\mathbf{t};z)\tau(\mathbf{t},\hat{\mathbf{t}})X(\mathbf{t}';-z)\tau(\mathbf{t}',\hat{\mathbf{t}}')\\
=&\res_z z^{-1}X(\hat{\mathbf{t}};-z^{-1})\tau(\mathbf{t},\hat{\mathbf{t}})X(\hat{\mathbf{t}}';z^{-1})\tau(\mathbf{t}',\hat{\mathbf{t}}'),
\end{align*}
which is equivalent to
\begin{align}
&\res_z z^{-1}X(\mathbf{t};z)\tau(\mathbf{t},\hat{\mathbf{t}})X(\mathbf{t}';-z)\tau(\mathbf{t}',\hat{\mathbf{t}}')\nn\\
=&\res_z z^{-1}X(\hat{\mathbf{t}};z)\tau(\mathbf{t},\hat{\mathbf{t}})X(\hat{\mathbf{t}}';-z)\tau(\mathbf{t}',\hat{\mathbf{t}}').
\label{bltau1}
\end{align}

Recall that in \cite{DJKM-KPtype, JM83}, Date, Jimbo, Kashiwara and
Miwa defined the two-component BKP hierarchy from a two-component
neutral free fermions realization of the basic representation of an
infinite-dimensional Lie algebra $\mfg_\infty$, which corresponds to
the Dynkin diagram of $D_{\infty}$ type \cite{Kac}. The tau function
of their hierarchy satisfies the bilinear equations \eqref{bltau1}
and defines  two wave functions as \eqref{wtau1},
\eqref{wtau2}, so the equations \eqref{ppt1},
\eqref{ppt2} give a representation of the two-component BKP hierarchy in terms of pseudo-differential operators.

\begin{rmk}
In \cite{Sh}, Shiota gave a Lax pair representation of the
two-component BKP hierarchy as follows. Let $\phi^{(\nu)}\ (\nu=0,
1)$ be the following pseudo-differential operators of the first type
\[\phi^{(\nu)}=1+\sum_{i\ge1}a^{(\nu)}_i D_\nu^{-i}\]
satisfying
$\left(\phi^{(\nu)}\right)^*=D_\nu\left(\phi^{(\nu)}\right)^{-1}D_\nu^{-1}$,
where $D_0, D_1$ are two commuting derivations. Let
\[
P^{(\nu)}=\phi^{(\nu)} D_\nu \left(\phi^{(\nu)}\right)^{-1},
\]
then the two-component BKP hierarchy can be defined as
\begin{equation}
\frac{\p \phi^{(\nu)}}{\p
t^{(\nu)}_k}=-\left(P^{(\nu)}\right)^k_-\phi^{(\nu)}, \quad \frac{\p
\phi^{(\nu)}}{\p
t^{(1-\nu)}_k}=\left(P^{(1-\nu)}\right)^k_+\left(\phi^{(\nu)}\right),
\quad k\in \Zop. \label{shiota}
\end{equation}
Here on the right hand side of the second equation it means the
action of the differential operator $\left(P^{(1-\nu)}\right)^k_+$
on the coefficients of $\phi^{(\nu)}$. It is easy to see that
$D_\nu=\frac{\p}{\p t^{(\nu)}_1}$. We identify $t^{(0)}_k=t_k$,
$t^{(1)}_k=\hat{t}_k$ henceforth.

Introduce the wave functions
\[w^{(\nu)}(\mathbf{t}, \hat{\mathbf{t}}; z^{(\nu)})=\phi^{(\nu)}e^{\xi^{(\nu)}},
\quad \xi^{(\nu)}=\xi(\mathbf{t}^{(\nu)};z^{(\nu)})
\]
with $\xi$ given in \eqref{xi}. The hierarchy \eqref{shiota} was
shown \cite{Sh} equivalent to the following bilinear equation
\begin{align}
&\res_{z^{(0)}}\big(z^{(0)}\big)^{-1}w^{(0)}(\mathbf{t}, \hat{\mathbf{t}}; z^{(0)})w^{(0)}(\mathbf{t}', \hat{\mathbf{t}}'; -z^{(0)}) \nn\\
=&\res_{z^{(1)}}\big(z^{(1)}\big)^{-1}w^{(1)}(\mathbf{t},
\hat{\mathbf{t}}; z^{(1)})w^{(1)}(\mathbf{t}', \hat{\mathbf{t}}';
-z^{(1)}). \label{shiota-bl}
\end{align}

By comparing the bilinear equations \eqref{shiota-bl} and
\eqref{bl2BKP}, it is easy to see that Shiota's wave functions are
related to ours by
\[w^{(0)}(\mathbf{t}, \hat{\mathbf{t}}; z)=w(\mathbf{t}, \hat{\mathbf{t}}; z),\quad
w^{(1)}(\mathbf{t}, \hat{\mathbf{t}}; z)=\hat{w}(\mathbf{t}, \hat{\mathbf{t}}; -z^{-1}),\]
from which one can obtain the relations between $a^{(0)}_i, a^{(1)}_i$ and $a_i, b_i$.
\end{rmk}

\subsection{Reductions of the two-component BKP hierarchy}

Given an integer $n\ge3$, the condition $P^{2n-2}=Q^2$ defines a
differential ideal of $\tilde{\mA}$, which is denoted by
$\mathcal{I}$. It is easy to see that this ideal is preserved by the
flows \eqref{ppt1}, \eqref{ppt2}, so we obtain a reduction of the
two-component BKP hierarchy.

Let $L=P^{2n-2}=Q^2$, then according to Lemma \ref{lempq} the
operator $L$ has the form \eqref{eq-LL}. Hence the algebra $\mA$
defined in Section 3.1 is isomorphic to $\tilde{\mA}/\mathcal{I}$,
and the reduced hierarchy is an integrable hierarchy over $\mA$. It
is easy to see that the derivatives of $L$ with respect to $t_k$,
$\hat{t}_k$ are exactly given by \eqref{dn1}, \eqref{dn2}. Namely
the hierarchy \eqref{dn1},\eqref{dn2} is the reduction of the
two-component BKP hierarchy under the condition $P^{2n-2}=Q^2$.

It can be shown that the condition $P^{2n-2}=Q^2$ reduces the
bilinear equations \eqref{bl2BKP} to the form
\begin{equation}
\res_z z^{(2n-2)j-1} w(\mathbf{t},\hat{\mathbf{t}}; z)w(\mathbf{t}',
\hat{\mathbf{t}}'; -z) =\res_z z^{-2 j-1}\hat{w}(\mathbf{t},
\hat{\mathbf{t}}; z) \hat{w}(\mathbf{t}',\hat{\mathbf{t}}'; -z)
\label{ble}
\end{equation}
with $j\ge 0$, and that conversely the equations \eqref{ble} impose
the constraint $P^{2n-2}=Q^2$ to the two-component BKP hierarchy.
Hence we establish the equivalence between the bilinear equations
\eqref{ble} and the hierarchy \eqref{dn1}, \eqref{dn2}. The proof is
lengthy and technical (c.f. the reduction from the KP hierarchy to
the Gelfand-Dickey hierarchies in \cite{Dickey}), so we omit the
details here. In terms of the tau function, the bilinear
equations \eqref{ble} can be expressed as
\begin{align} \label{bltau}
&\res_z z^{(2n-2)j-1}X(\mathbf{t};z)\tau(\mathbf{t},\hat{\mathbf{t}})X(\mathbf{t}';-z)\tau(\mathbf{t}',\hat{\mathbf{t}}')\nn\\
=&\res_z
z^{2j-1}X(\hat{\mathbf{t}};z)\tau(\mathbf{t},\hat{\mathbf{t}})X(\hat{\mathbf{t}}';-z)\tau(\mathbf{t}',\hat{\mathbf{t}}'),
\quad j\ge0.
\end{align}
Note that these bilinear equations are precisely the ones obtained
from the $(2n-2,2)$-reduction of the two-component BKP hierarchy
\cite{DJKM-reduce, JM83}.

From the definition \eqref{tauT} and \eqref{tau-2} of the tau functions $\hat\tau$ and $\tau$ it follows that they are related by
\begin{equation}\label{tauhtau}
\tau^2=\hat{\tau}.
\end{equation}

\section{Conclusion}

We represent the full Drinfeld-Sokolov hierarchy of $D_n$ type into
 Lax equations of pseudo-differential
operators, which is analogous to the Gelfand-Dickey hierarchies. We
also give a Lax pair representation for the two-component BKP
hierarchy, and show that the Drinfeld-Sokolov hierarchy of $D_n$ type
is the $(2n-2,2)$-reduction of the two-component BKP hierarchy. The key step
in our approach is to introduce the concept of pseudo-differential operators of the second
type, which are defined over a topologically complete differential algebra, so that they may
contain infinitely many terms with positive power of the derivation $D$.

Our Lax pair representations of the Drinfeld-Sokolov hierarchy of $D_n$ type and the two-component BKP
hierarchy are convenient for further studies. In a subsequent publication \cite{BKP2},
we will show that the two-component BKP hierarchy carries a bihamiltonian structure, which is expected to correspond to an infinite-dimensional Frobenius
manifold (c.f. \cite{CDM}).

Note that the bilinear equation \eqref{bltau1} corresponds to the basic representation of the affine Lie algebra $D_{\infty}'$ in the notion of
\cite{JM83}. It is shown in \cite{tv} that the $(2n-2,2)$-reduction \eqref{bltau} corresponds to the basic representation of the affine Lie algebra $D_n^{(1)}$.
Then according to \cite{Kac, KW}, the bilinear equation \eqref{bltau} is equivalent to the Kac-Wakimoto hierarchy constructed from the
principal vertex operator realization of the basic representation of the affine Lie algebra $D_n^{(1)}$ \cite{KW}.
By comparing the boson-fermion correspondences, one can obtain the relation between the time variables $\mathbf{t}, \hat{\mathbf{t}}$
of the Drinfeld-Sokolov hierarchy of $D_n$ type (or the Date-Jimbo-Kashiwara-Miwa hierarchy) and the time variables
$s_j\ (j\in E_+)$ of the the Kac-Wakimoto hierarchy
\[t_k=\sqrt{2}\,s_k, \quad \hat{t}_k=\sqrt{2n-2}\,s_{k(n-1)'}.\]
In \cite{GM}, Givental and Milanov proved that the total descendant potential for semisimple Frobenius manifolds associated to a simple
singularity satisfies a certain hierarchy of Hirota bilinear/quadratic equations, see also \cite{Gi, Gi-GW, Gi-A}. Such a hierarchy of
bilinear equation is shown to be equivalent to the corresponding Kac-Wakimoto hierarchy constructed from the principal vertex operator
realization of the basic representation of the untwisted affine Lie algebra \cite{GM, Wu, FGM}.
So we arrive at the following result.
\begin{thm}\label{thm-id}
Up to a rescaling of the flows, the following integrable hierarchies
are equivalent:
\begin{itemize}
\item[i)] the hierarchy \eqref{dn1}, \eqref{dn2};
\item[ii)] the Drinfeld-Sokolov hierarchy associated to $D_n^{(1)}$ and the $c_0$ vertex of its Dynkin diagram;
\item[iii)] the Date-Jimbo-Kashiwara-Miwa hierarchy constructed from the basic representation of the affine Lie
algebra $D_n^{(1)}$;
\item[iv)] the Kac-Wakimoto hierarchy corresponding
to the principal vertex operator realization of the basic
representation of the affine Lie algebra $D_n^{(1)}$;
\item[v)] the Givental-Milanov hierarchy for the simple singularity of $D_n$ type.
\end{itemize}
\end{thm}

\begin{rmk}
The equivalence between the hierarchies ii) and iv) was also
contained in a general result obtained by Hollowood and Miramontes
in \cite{HM}.

Note that the bihamiltonian structure \eqref{poi1}, \eqref{poi2} is of topological type \cite{DLZ-1,DZ, DLZ},
its leading term comes from the Frobenius manifold associated to the Coxeter group of $D_n$ type.
In \cite{DZ} a hierarchy of dispersionless bihamiltonian integrable systems
is associated to any semisimple Frobenius manifold, such an integrable hierarchy is called the Principal Hierarchy. It is also shown that there is a so called topological deformation of the Principal Hierarchy which satisfies the condition that its Virasoro symmetries can be represented by the action of some linear operators, called the Virasoro operators, on the tau function of the hierarchy.
We expect that the Drinfeld-Sokolov hierarchy associated to $D_n^{(1)}$ and the $c_0$ vertex of its Dynkin diagram
 coincides, after a rescaling of the time variables, with the topological deformation of the Principal Hierarchy of the Frobenius manifold that is associated to the Coxeter group of type
$D_n$. We will investigate this aspect of the hierarchy in
a subsequent publication.
\end{rmk}

\vskip 1ex \noindent{\bf Acknowledgments.} The authors thank Boris
Dubrovin for his interest in this work and for his advices, they also thank for the
hospitality of  SISSA where part of the work was done. This work is
partially supported by the National Basic Research Program of China
(973 Program) No.2007CB814800, the NSFC No.10631050 and No.10801084.


\begin{thebibliography}{99}
\bibitem{CDM}  Carlet, G.; Dubrovin, B.; Mertens L.P. Infinite-dimensional Frobenius manifolds for 2+1 integrable
systems, preprint arXiv: math.ph/0902.1245v1.

\bibitem{DJKM-reduce}Date, E.; Jimbo, M.; Kashiwara, M.; Miwa, T.
Transformation groups for soliton equations. Euclidean Lie algebras
and reduction of the KP hierarchy. Publ. Res. Inst. Math. Sci. 18
(1982), no.3, 1077--1110.
\bibitem{DJKM-KPtype} Date, E.; Jimbo, M.; Kashiwara, M.; Miwa, T.
Transformation groups for soliton equations. IV. A new hierarchy of
soliton equations of KP-type. Phys. D 4 (1981/82), no.3, 343--365.
\bibitem{DKJM-KPBKP}
Date, E; Kashiwara, M; Jimbo, M; Miwa, T. Transformation groups for
soliton equations. Nonlinear integrable systems---classical theory
and quantum theory (Kyoto, 1981), 39--119, World Sci. Publishing,
Singapore, 1983.
\bibitem{Dickey} Dickey, L.A. Soliton equations and Hamiltonian systems. Second edition.
Advanced Series in Mathematical Physics, 26. World Scientific Publishing Co., Inc., River Edge, NJ, 2003.

\bibitem{DS} Drinfeld, V.G.; Sokolov, V.V.
Lie algebras and equations of Korteweg-de Vries type. (Russian)
Current problems in mathematics, Vol. 24, 81--180, Itogi Nauki i
Tekhniki, Akad. Nauk SSSR, Vsesoyuz. Inst. Nauchn. i Tekhn. Inform.,
Moscow, 1984.
\bibitem{Du} Dubrovin, B. Geometry of $2$D topological field theories.
Integrable systems and quantum groups (Montecatini Terme, 1993),
120--348, Lecture Notes in Math., 1620, Springer, Berlin, 1996.
\bibitem{DLZ-1}Dubrovin, B.; Liu, S.Q.; Zhang Y. On Hamiltonian
perturbations of hyperbolic systems of conservation laws I:
quasi-triviality of bi-Hamiltonian perturbations, Commun. Pure and
Appl. Math. 59 (2006), 559-615.
\bibitem{DLZ} Dubrovin, B.; Liu, S.Q.; Zhang, Y. Frobenius manifolds and central invariants for the Drinfeld-Sokolov biHamiltonian structures. (English summary)
Adv. Math. 219 (2008), no.3, 780--837.
\bibitem{DZ} Dubrovin, B.; Zhang, Y. Normal forms of integrable PDEs,
Frobenius manifolds and Gromov-Witten invariants, preprint arXiv:
math.DG/0108160.
\bibitem{EF} Enriquez, B.; Frenkel, E. Equivalence of two approaches to integrable hierarchies of KdV type. Comm. Math. Phys. 185 (1997),  211--230.
\bibitem{FSZ}Faber, C.; Shadrin, S.; Zvonkine, D.
Tautological relations and the r-spin Witten conjecture, preprint
arXiv:math.AG/0612510.
\bibitem{FJR} Fan, H.; Jarvis, T.J.; Ruan, Y. The Witten equation, mirror symmetry and quantum singularity
theory, preprint arXiv: math.AG/0712.4021v3.
\bibitem{FF} Feigin, B.; Frenkel, E. Kac-Moody groups and integrability of soliton equations. Invent. Math. 120 (1995),  379--408.
\bibitem{HM-2}  Ferreira, L.A.; Miramontes, J.L.; S¨¢nchez Guill\'en, J. Tau-functions and dressing transformations for zero-curvature affine integrable equations. J. Math. Phys. 38 (1997), no.2, 882--901.
\bibitem{FGM} Frenkel, E.; Givental, A.; Milanov, T. Soliton equations, vertex operators, and simple
singularities, preprint arXiv:math.QA/0909.4032, 2009.

\bibitem{GD76} Gelfand, I.M.; Dikii, L. A.
Fractional powers of operators, and Hamiltonian systems. (Russian)
Funkcional. Anal. i Prilozen. 10 (1976), no.4, 13--29.
\bibitem{Gi} Givental, A. Semi-simple Frobenius structures at higher genus. International Mathematics
Research Notices 2001, no.23 (2001): 1265-1286.
\bibitem{Gi-GW} Givental, A. Gromov-Witten invariants and quantization of quadratic Hamiltonians.
Dedicated to the memory of I. G. Petrovskii on the occasion of his
100th anniversary. Mosc. Math. J. 1 (2001), no.4, 551--568, 645.
\bibitem{Gi-A} Givental, A. $A_{n-1}$-singularities and $n$KdV hierarchies. Dedicated to Vladimir I. Arnold on the occasion of his 65th birthday.
Mosc. Math. J. 3 (2003), no.2, 475--505, 743.
\bibitem{GM} Givental, A.; Milanov, T. Simple singularities and
integrable hierarchies. The breadth of symplectic and Poisson
geometry, 173--201, Progr. Math., 232, Birkh\"{a}user Boston,
Boston, MA, 2005.
\bibitem{GHM}  de Groot, M.F.; Hollowood, T.J.; Miramontes, J.L.
Generalized Drinfeld-Sokolov hierarchies. Comm. Math. Phys. 145
(1992), no.1, 57--84.
\bibitem{HM}  Hollowood, T.J.; Miramontes, J.L.
Tau-functions and generalized integrable hierarchies. Comm. Math.
Phys. 157 (1993), no.1, 99--117.\bibitem{JM83}  Jimbo, M.; Miwa, T.
Solitons and infinite-dimensional Lie algebras. Publ. Res. Inst.
Math. Sci. 19 (1983), no.3, 943--1001.
\bibitem{Kac} Kac, V.G. Infinite-dimensional Lie algebras. Third edition. Cambridge University Press, Cambridge, 1990, RI, 1989.
\bibitem{KW} Kac, V.G.; Wakimoto, M. Exceptional hierarchies of soliton equations.
Theta functions---Bowdoin 1987, Part 1 (Brunswick, ME, 1987),
191--237, Proc. Sympos. Pure Math., 49, Part 1, Amer. Math. Soc.,
\bibitem{Kon} Kontsevich, M. Intersection theory on the moduli space of curves and the matrix Airy function.
Comm. Math. Phys. 147 (1992), no.1, 1--23.
\bibitem{tv} ten Kroode, F.; van de Leur, J. Bosonic and fermionic realizations of the affine algebra $\widehat{\rm so}_{2n}$. Comm. Algebra 20 (1992), no.11, 3119--3162.
\bibitem{Sh} Shiota, T. Prym varieties and soliton equations.
Infinite-dimensional Lie algebras and groups (Luminy-Marseille,
1988), 407--448, Adv. Ser. Math. Phys., 7, World Sci. Publ.,
Teaneck, NJ, 1989.
\bibitem{Ta} Takasaki, K. Integrable hierarchy underlying topological Landau-Ginzburg models of $D$-type. Lett. Math. Phys. 29 (1993), no.2, 111--121.
\bibitem{Wi} Wilson, G. The modified Lax and two-dimensional Toda lattice equations associated with simple Lie algebras. Ergodic Theory Dynamical Systems 1 (1981), no.3, 361--380 (1982).
\bibitem{Witten1} Witten, E. Two-dimensional gauge theories revisited. J. Geom. Phys. 9 (1992), no.4, 303--368.
\bibitem{Wu} Wu, C.Z. A Remark on Kac-Wakimoto Hierarchies
of D-type. J. Phys. A: Math. Theor. 43 (2010), 035201.
\bibitem{BKP2} Wu, C.Z.; Xu, D. Bihamiltonian structure of the two-componet BKP hierarchy, in preparation.
\end{thebibliography}
\end{document}